\documentclass[iop,revtex4]{emulateapj}
\usepackage{lscape}
\usepackage{csquotes}
\usepackage{amssymb}
\usepackage{amsmath}
\usepackage{mathtools}
\usepackage{booktabs}

\usepackage[breaklinks,colorlinks,urlcolor=blue,citecolor=blue,linkcolor=blue]{hyperref}
\shortauthors{Chou et al.}

\begin{document}
\submitted{Published in the Astrophysical Journal}

\title{Protostar L1455 IRS1: Rotating Disk Connecting to Filamentary Network}

\author{Hsuan-Gu Chou\altaffilmark{1,2}, Hsi-Wei Yen\altaffilmark{1}, Patrick M. Koch\altaffilmark{1} and St\'ephane Guilloteau\altaffilmark{3}}
\altaffiltext{1}{Academia Sinica Institute of Astronomy and Astrophysics, P.O. Box 23-141, Taipei 10617, Taiwan}
\email{hgchou@asiaa.sinica.edu.tw}
\altaffiltext{2}{Physics Department, National Taiwan University.
 No. 1, Sec. 4, Roosevelt Road, Taipei, 10617 Taiwan (R.O.C)}
 \altaffiltext{3}{Universit\'e de Bordeaux, Observatoire Aquitain des Sciences de l'Univers, CNRS, UMR 5804, Laboratoire d'Astrophysique de Bordeaux, 2 rue de l'Observatoire, BP 89, F-33271 Floirac Cedex, France}
 
\begin{abstract}
We conducted IRAM-30m C$^{18}$O (2--1) and SMA 1.3 mm continuum, $^{12}$CO (2--1), and C$^{18}$O (2--1) observations toward the Class 0/I protostar L1455 IRS1 in Perseus. The IRAM-30m C$^{18}$O results show IRS1 in a dense 0.05 pc core with a mass of 0.54 $M_\sun$, connecting to a filamentary structure.
Inside the dense core, compact components of 350 AU and 1500 AU are detected in the SMA 1.3 mm continuum and C$^{18}$O, with a velocity gradient in the latter one perpendicular to a bipolar outflow in $^{12}$CO, likely tracing a rotational motion.
We measure a rotational velocity profile $\propto r^{-0.75}$ that becomes shallower at a turning radius of $\sim$200 AU which is approximately the radius of the 1.3 mm continuum component.
These results hint the presence of a Keplerian disk with a radius $<$200 AU around L1455 IRS1 with a protostellar mass of about 0.28 $M_\sun$.
We derive a core rotation that is about one order of magnitude faster than expected. A significant velocity gradient along a filament towards IRS1 indicates that this filament is dynamically important, providing a gas reservoir and possibly responsible for the faster-than-average core rotation.
Previous polarimetric observations show 
a magnetic field aligned with the outflow axis and perpendicular to 
the associated filament on a 0.1~pc scale, while on the inner 1000 AU scale, the field becomes perpendicular to the outflow axis.
This change in magnetic field orientations is consistent with our 
estimated increase in rotational energy from large to small scales
that overcomes the magnetic field energy, wrapping the field lines and 
aligning them with the disk velocity gradient.
These results are discussed in the context of the interplay between filament, magnetic field, and gas kinematics from large to small scale.
Possible emerging trends are explored with a sample of 8 Class 0/I protostars. 
\end{abstract}


\section{Introduction}\label{sec:intro}
The study of star-forming regions is one of the most 
eminent fields in astronomy. Enhanced instrumental capabilities are now enabling us 
to investigate regions which were impossible to detect or resolve before. In particular, the advent of (sub-)millimeter detectors is providing us with opportunities to plumb more deeply the mystery of star formation (e.g. \citet{johnstone2004,kirk2005}). 
To a large extent this is because dust
usually becomes optically thin at the longer (sub-)millimeter wavelengths. Hence, estimates for total masses of molecular clouds, cores and envelopes become possible as the emission traces the entire structures \citep{enoch2006}. Molecular clouds are the birthplaces of stars (e.g. \citet{wilson1996}). 
Numerous clouds are found in the local Milky Way. Among them, the Perseus molecular cloud has been studied for decades and is a proven star-forming factory.

The Perseus cloud extends over an area of about hundred pc$^2$, elongated from north-east to south-west with a length of 25 pc (e.g. \citet{ungerechts1985}). With a total mass around $10^4M_\odot$ \citep{sancisi1974,evans2009,sadavoy2010}, it is filled with protostellar and star-forming clusters, such as NGC 1333, L1448, and L1455. The very first molecular line surveys of the Perseus OB2 association (containing massive stars) in $^{12}$CO \citep{sargent1979} and $^{12}$CO, $^{13}$CO, C$^{18}$O \citep{bachiller19861}, suggested a complex structure of the Perseus molecular cloud, which is believed to be influenced by and thus, physically related to the OB2 association (for a summary of the Perseus cloud, see e.g., \citet{bally2008}).
The distance to the Perseus cloud has been a long-lasting controversy in the literature. Measurements toward the eastern and western ends of the cloud reveal significant discrepancies. For example, \citet{cernis1993} suggests a distance of 220 pc to the western part NGC 1333, while the \textit{Hipparcos} parallax census shows  320~pc for the eastern IC 348 \citep{de1999}. High-accuracy maser parallax measurements give $\sim$232~pc for L1448. Besides, a clear velocity gradient extending over $\sim$50~pc in length from 3 km s$^{-1}$ in the western edge to 10 km s$^{-1}$ in the eastern end \citep{sargent1979,ungerechts1985,padoan1999,sun2006} is thought to result from multiple structures overlapping along the line of sight \citep{ridge2006}, which adds further uncertainty to determining a unified distance. Recent studies (e.g. \citet{enoch2006,jorgensen2006,ridge2006,evans2009,arce2010,curtis20101,sadavoy2010}) adopt a distance of 250 pc, which we also assume for 
the present work.
At this distance, $1\arcsec$ corresponds to about 250~AU. 

\subsection{L1455 Complex}

\begin{figure*}[!ht]
\centering
\includegraphics[width=\textwidth]{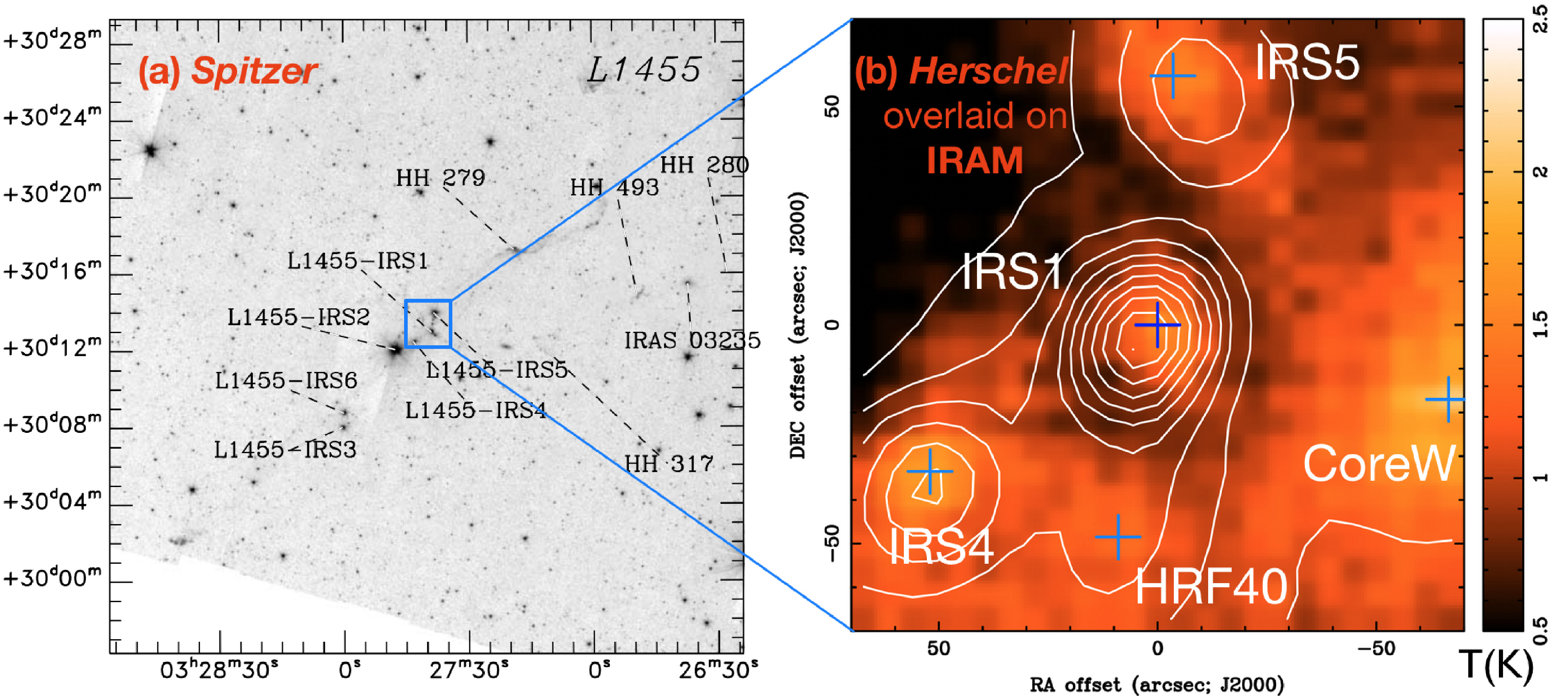}
\caption{(a) L1455 \textit{Spitzer} 4.5$\mu m$ image retrieved from \citet{jorgensen2006}. (b) \textit{Herschel} 350$\mu m$ image (contours, starting at 10\% of peak value in steps of 10\%) overlaid on our IRAM-30m C$^{18}$O integrated intensity map (color wedge). The \textit{Herschel} map is reproduced from archival data \citep{andre2010}.
}
\label{fig:fil}
\end{figure*}

Within the Perseus cloud, besides the two young clusters IC 348 and NGC 1333, there are mainly four small regions which are spawning new stars: B1, B5, L1448, and L1455 (e.g., \citet{young2015}). The L1455 cloud, being an intermediate mass case, possesses a suitable number of possibly interacting and forming stars (e.g. \citet{curtis20102}). It, thus, serves as an ideal testbed to study star formation in a cluster. Furthermore, \citet{jorgensen2008} hinted that, in regions like L1448 and L1455, the deeply embedded protostars are likely in their earliest stages of evolution. Therefore, examining these regions is a key to understanding star formation in its early phase.

The L1455 region \citep{lynds1962} encloses an area of about 7.6 pc$^2$, centered approximately at $\alpha$(J2000)$=3^h28^m00^s$, $\delta$(J2000)$=30^\circ20^\prime00$. It contains 11 young stellar objects (YSOs), with the majority of them being in their early stages (Class 0/$\mathrm{I}$) \citep{young2015}. The first three protostars discovered in the \textit{IRAS}\footnote{Infrared Astronomical Satellite} catalogs are L1455 IRS1, L1455 IRS2, and L1455 IRS3 \citep{IRAS}. Since then, more were found in the proximity of them. Specifically, \citet{jorgensen2006} name three more protostars, L1455 IRS4, L1455 IRS5 and L1455 IRS6, that are all clustered inside a small $\sim 1$ pc$^2$ region.
Unfortunately, this area where more than half of the YSOs reside, as well as the individual cores were seldom investigated in detail. In particular, the densest region in L1455, encompassing a roughly 0.02 pc$^2$ box centering on L1455 IRS1 has never been studied with scrutiny.

L1455 IRS2 \citep{juan1993} (a.k.a. RNO 15 \citep{cohen1980} or PP9 \citep{tapia1997}) is probably the very first discovered YSO in the region. \citet{frerking1982,goldsmith1984,levreault1988} provide early CO images of outflows that  initiated the discussion of the extended outflows observed around L1455 IRS2. An early NH$_3$ survey also confirms these outflows \citep{anglada1989}. Several Herbig-Halo objects were also identified via H$\alpha$ and $[\text{S}\mathrm{II}]$ \citep{bally1997}. \citet{davis19971} use both $[\text{S}\mathrm{II}]$ and $^{12}$CO lines to analyze these outflows, finding a dominant northwest-to-southeast elongated outflow overlapping a smaller one perpendicular to it. Although a number of studies ascribe the origin of the first (dominating) one to L1455 IRS2 and/or L1455 IRS5 \citep{davis2008,curtis20102,walker2014}, others are in favor of attributing it to L1455 IRS4, which sits approximately in the center of the outflow \citep{jorgensen2006,simon2009,arce2010}. 
While there might be multiple outflows along the northwest-southeast direction,
\citet{jorgensen2006} exclude L1455 IRS2 and L1455 IRS5 as candidate driving sources based
on the evolutionary stages of the surrounding protostars. This is further supported by
\citet{wu2007} and NH$_3$ imaging \citep{sepulveda2011}. 

Figure \ref{fig:fil}(a) shows the L1455 region surveyed by \citet{jorgensen2006} with 
\textit{Spitzer}. Figure \ref{fig:fil}(b) depicts the \textit{Herschel} 350$\mu m$ detection  
overlapping with our IRAM-30m C$^{18}$O integrated intensity map (Section \ref{sec:obs_iram}). Tracing back to \textit{Spitzer} results, \citet{hatchell2005} show that the L1455 cloud has a filamentary appearance with a southeast-northwest axis, coinciding with the complex dominant outflow mentioned above. Since this outflow is covering several protostars on its path, the debate on its source is still unsettled. As early as \citet{juan1993}, L1455 IRS1 was viewed as the driving source of the second outflow in this region. Recently, a survey by CARMA\footnote{Combined Array for Research in Millimeter-wave Astronomy} verified this, though with a slightly different outflow axis position angle \citep{hull2014}.

Since the \textit{Herschel} Gould Belt survey, filaments are ubiquitously observed in star-forming regions (e.g., \citet{andre2010}). Alongside high-mass star-forming regions, low-mass ones are no exception \citep{di2012}. Recent studies, using observations with higher angular resolutions and kinematic information, further confirm the significance of filaments in such regions (e.g. \citet{hacar2011,tafalla2015}). The detailed mechanisms according to which pre-stellar cores form from filaments, however, are still obscure (e.g., \citet{ward2007,di2007}) and, although widely seen, these hierarchical structures (from clouds to filaments to cores) in clustered star-forming regions remain mysterious \citep{bergin2007}. Therefore, our goal of the present paper is to provide an in-depth study addressing star formation in a complex cloud in a  filamentary environment.

\subsection{L1455 IRS1}
As alluded to in the previous paragraphs, the L1455 region possess filaments as well. In this paper, we focus on one specific protostar in this complex cloud, L1455 IRS1. It is one of the brightest protostars ($L_{\text{bol}}=3.6L_\odot$) in the L1455 region \citep{dunham2013}. As a protostar in its early stage, it has an ambiguous classification throughout the literature. The discrimination between Class 0 and Class $\mathrm{I}$ protostars, from an observational point of view, is non-trivial (e.g., \citet{dunham2014}). Thus, we follow the suggestion of \citet{young2015}, where L1455 IRS1 is assigned a Class $0/\mathrm{I}$. 

Abundant archival data are enabling us to examine L1455 IRS1 from different perspectives.
Dust continuum polarimetry data from SCUPOL and TADPOL are also available. The SCUPOL legacy \citep{matthews2009} provides large-scale information of the magnetic field in the densest sub-region, similar in size to our IRAM-30m single-dish region in L1455. The TADPOL survey using CARMA \citep{hull2014} yields high-resolution polarization data specific to L1455 IRS1. We obtained new data with the SMA 
and the IRAM-30m. Section \ref{sec:rs}
presents our images of L1455 IRS1 and its surrounding. 
Section \ref{sec:ana} details our analyses supporting the existence of a disk around L1455 IRS1. Finally, we discuss the larger scale connection in the filamentary network together with magnetic field implications in section \ref{sec:dis}, and we conclude with possible trends by looking at a sample of 8 sources.

\section{Observations}\label{sec:obs}
\subsection{SMA}\label{sec:smaobs}
The SMA observations of L1455 IRS1 were made in the subcompact configuration with seven antennas on 2014, July 24 and in the compact-north configuration with eight antennas on 2014, September 15. 
Details of the SMA are described in \citet{sma}. 
The 230 GHz receiver with a bandwidth of 2 GHz per sideband was adopted for our observations. 
The 1.3 mm continuum, and the $^{12}$CO (2--1) and C$^{18}$O (2--1) lines were observed simultaneously. 
2048 channels were assigned to a chunk with a bandwidth of 104 MHz for the C$^{18}$O line, 
and 512 channels for the $^{12}$CO line, 
corresponding to velocity resolutions of 0.07 km s$^{-1}$ and 0.26 km s$^{-1}$, respectively.
Combing the two array configurations, the projected baseline lengths range from $\sim$5 to $\sim$100 k$\lambda$. 

For both observing nights 3c454.3 and Uranus were observed as bandpass and flux calibrators. 
In the observation on July 24, 
the gain calibrators were 3c84 (11.1 Jy) and 3c111 (1.7 Jy). 
The atmospheric opacity at 225 GHz was 0.14--0.16, 
and the typical system temperature was 100--200 K. 
On September 15, 
the gain calibrator was 3c84 (11.5 Jy) with 
an atmospheric opacity at 225 GHz of 0.18--0.26, 
and a system temperature ranging from 120 to 400 K. 
The MIR software package \citep{mir} was used to calibrate the data. 
The calibrated visibility data obtained with the compact and subcompact configurations were Fourier-transformed together and CLEANed with MIRIAD \citep{miriad}. 
Images are made to be 512 pixels$\times$512 pixels with a cell size of 0\farcs2$\times$0\farcs2, resulting in a map size of $\sim1\farcm7\times 1\farcm7$.
Our displayed maps are limited to central sections with significant detections.
Sizes of the beams and noise levels of our images are summarized in Table \ref{tb:obs}. 

\begin{deluxetable*}{lccc}[!ht]
\tablecaption{Summary of L1455 IRS1 Observations\label{tb:obs}}

\tablehead{ & \multicolumn{3}{c}{Value}\\ \cline{2-4}
           \colhead{Parameter} & \colhead{SMA} & \colhead{Combined} & \colhead{IRAM-30m}} 
\startdata
Synthesized beam / resolution FWHM (line) & 4\farcs54$\times$2\farcs74 (P.A.$=71^\circ$) & 6\farcs1$\times$3\farcs5 (P.A.$=75^\circ$) & 11\farcs8$\times$11\farcs8 \\
Synthesized beam FWHM (continuum) & 3\farcs06$\times$1\farcs51 (P.A.$=74^\circ$) & -- & -- \\
Velocity resolution (C$^{18}$O) & 0.069 km s$^{-1}$ & 0.07 km s$^{-1}$ & 0.027 km s$^{-1}$ \\
Primary beam size / FoV (C$^{18}$O) & 50\farcs4 & -- & 72\farcs8  \\
Primary beam size ($^{12}$CO) & 48\farcs0 & -- & -- \\
rms noise level (continuum) & 1.55 mJy beam$^{-1}$ & -- & -- \\
rms noise level (C$^{18}$O) (single channel)& 161 mJy beam$^{-1}$ & 105 mJy beam$^{-1}$ & 0.24 K \\
rms noise level ($^{12}$CO) (single channel) & 144 mJy beam$^{-1}$ & -- & -- \\
\enddata
\end{deluxetable*}

\subsection{IRAM-30m} \label{sec:obs_iram}
The IRAM-30m C$^{18}$O (2-1) observations toward L1455 IRS1 were conducted on 2014, Sepember 3, 4, and 6. The IRAM-30m telescope is described in detail in \citet{iram}. 
For our observations, the HERA heterodyne receiver with a 3-by-3 dual polarization pixel pattern was adopted. The VESPA autocorrelator served as the backend. It was set to have a spectral resolution of 20 kHz over a bandwidth of 20 MHz, resulting in a velocity resolution of $\sim$0.03 km s$^{-1}$ for the C$^{18}$O (2-1) line.
The observations were conducted in the position-switching on-the-fly (OTF) mode. 
The beam pattern of the HERA receiver was rotated by 9\arcdeg5 
in order to oversample.
The length and spacing of the OTF scans were set to map a 2$\arcmin$-by-2$\arcmin$ area centered on L1455 IRS1 in a homogeneous sampling. 
Pixel 4 and 9 of HERA2 were not functioning, and their data were excluded. 
During the observations, the precipitable water vapor (PWV) ranged from 2 to 9 mm, and the system temperature from 400 to 750 K. 
We performed baseline calibration and generated an image cube using the Continuum and Line Analysis Single-dish Software (CLASS). 
The angular resolution of this image cube is $\sim$11\farcs8.  
The main beam efficiency of the IRAM-30m telescope was 0.65. 
We scale the observed antenna temperature in our image cube to the main beam temperature using this efficiency.

\subsection{Combining SMA and IRAM-30m data}
To recover the missing flux in the SMA maps ($\sim$70\% as shown below) and to explore possibly relevant structures between the scales probed by the SMA and IRAM-30m observations, 
we combine our SMA and IRAM-30m C$^{18}$O (2--1) data. 

The size of the combined map is given by the largest detectable scales, i.e., by IRAM-30m data. The addition of the SMA data can refine and sharpen structures where the SMA has detection.
Generally, this will lead to more emphasized features with more details.
We follow the method described in \citet{yen2011}.
We further test our combining process to demonstrate that 
no artificial structures are created (Appendix \ref{sec:test}). 
The combined image cube is generated at a velocity resolution of 0.07 km s$^{-1}$, and has an angular resolution of 6\farcs1$\times$3\farcs5 and a noise level of 105 mJy Beam$^{-1}$.

\section{Results}\label{sec:rs}
\subsection{SMA}
\subsubsection{1.3 mm  Continuum Emission}\label{sec:con}
\begin{figure*}[!ht]
\centering
\includegraphics[width=\textwidth]{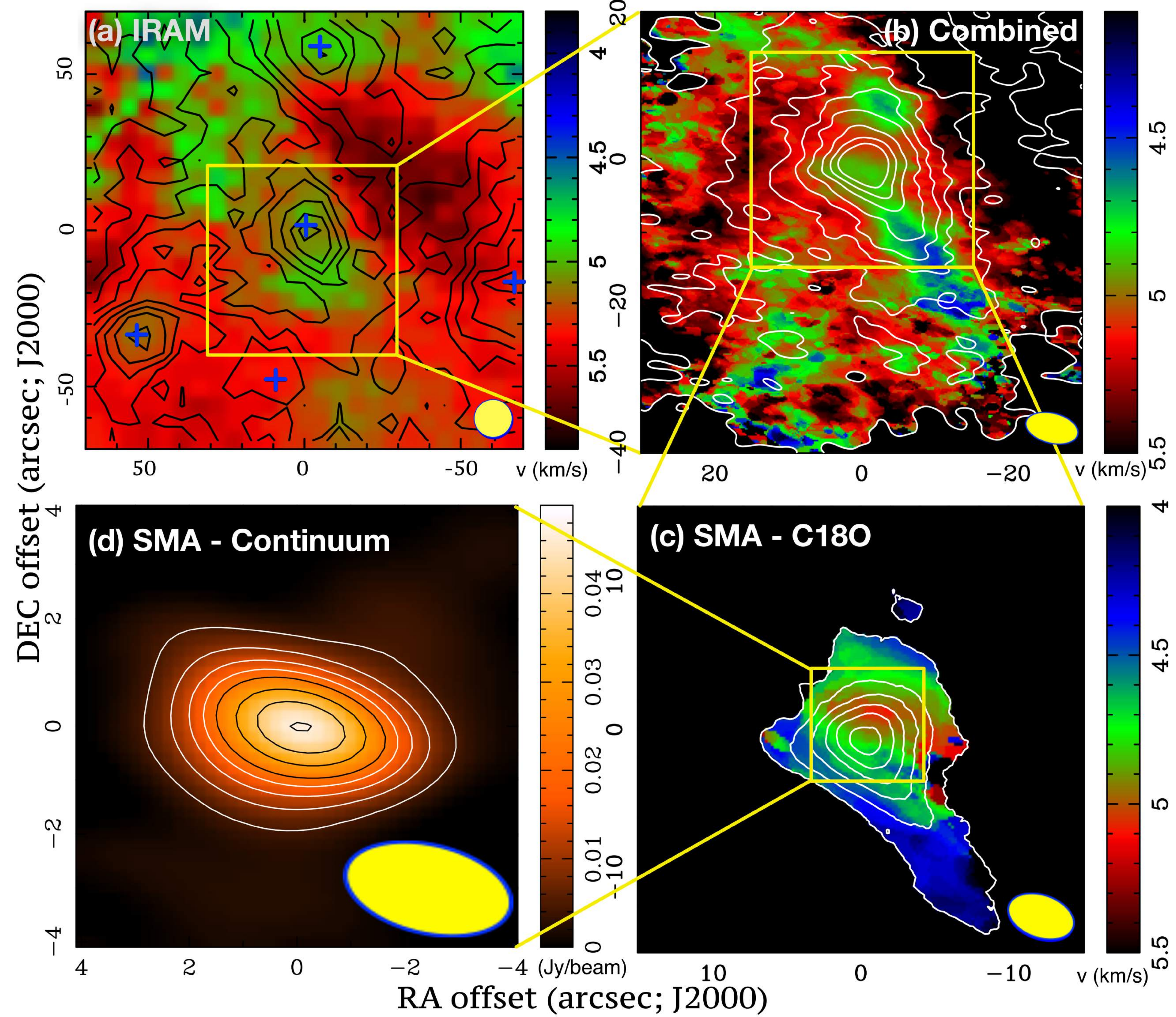}
\caption{(a) IRAM-30m C$^{18}$O (2--1) single-dish moment 0 map (contour) overlaid on the moment 1 map (color) around L1455 IRS1. (b)
As in (a) but for the SMA-IRAM-combined data, zooming in onto the yellow square in (a). (c) As in (a) but for SMA C$^{18}$O (2--1).
(d) SMA 1.3mm continuum map.
A filled ellipse in the bottom right corner in each panel denotes the beam size. Crosses show the protostellar positions. Contour levels are from 5$\sigma$ in steps of 5$\sigma$ in (a) and from 3$\sigma$ in steps of 3$\sigma$ in (c), where 1$\sigma$ is 0.034 K km s$^{-1}$ in (a) and 100 mJy beam$^{-1}$ km s$^{-1}$ in (c), respectively. Contours start from 3$\sigma$ in steps of 3$\sigma$ in (b), where 1$\sigma$ is 92 mJy beam$^{-1}$ km s$^{-1}$, and contours are 3$\sigma$ to 15$\sigma$ in steps of 3$\sigma$, and then in 5$\sigma$ in (d), where 1$\sigma$ is 1.55 mJy beam$^{-1}$.}
\label{fig:m0}
\end{figure*}

Figure \ref{fig:m0}(d) shows the observed 1.3 mm continuum image of L1455 IRS1. By fitting a two-dimensional Gaussian with the MIRIAD fitting program \textit{imfit}, we obtain a peak position of $\alpha$(J2000) = 03$^h$27$^m$39$^s$.1, $\delta$(J2000) = 30$^\circ$13$^\prime 03\farcs0$. In the present paper, we adopt this as the position of the protostar.
This is consistent with the previous 1.1 mm continuum result obtained by the CSO \citep{enoch2009}. The deconvolved size, position angle, and total flux are estimated to be $1\farcs4 \times 0\farcs8$ (350 $\times$ 200 AU), 129$^\circ$, and 61 mJy, respectively. 

We estimate the dust mass ($\equiv M_\text{dust}$) as
\begin{align}
M_\text{dust} = \frac{F_\nu D^2}{\kappa_\text{230GHz}B(T_\text{dust})}
\end{align}
where $F_\nu$ is the total flux, $D$ is the distance to the source, $T_\text{dust}$ is the dust temperature taken to be 50K which is a typical disk temperature at a radius of a few hundred AU (e.g., \citet{pietu2007}). $B(T_\text{dust})$ is the Planck function at the temperature $T_\text{dust}$. On the assumption that the frequency ($\nu$) dependence of the dust mass opacity ($\kappa_\nu$) is $\kappa_\nu = 0.1 \times (\frac{\nu}{10^{12}})^\beta$ \citep{beckwith1990}, the mass opacity at 1.3 mm (230 GHz) is 0.023 cm$^2$ g$^{-1}$ with $\beta = 1.0$ (e.g., \citet{jorgensen20072}) and a gas-to-dust mass ratio of 100. The total mass is estimated to be $0.011 M_\odot$.

\subsubsection{$^{12}$CO (2--1) Emission}\label{sec:12co}
\begin{figure}[!ht]
\centering
\includegraphics[width=0.49\textwidth]{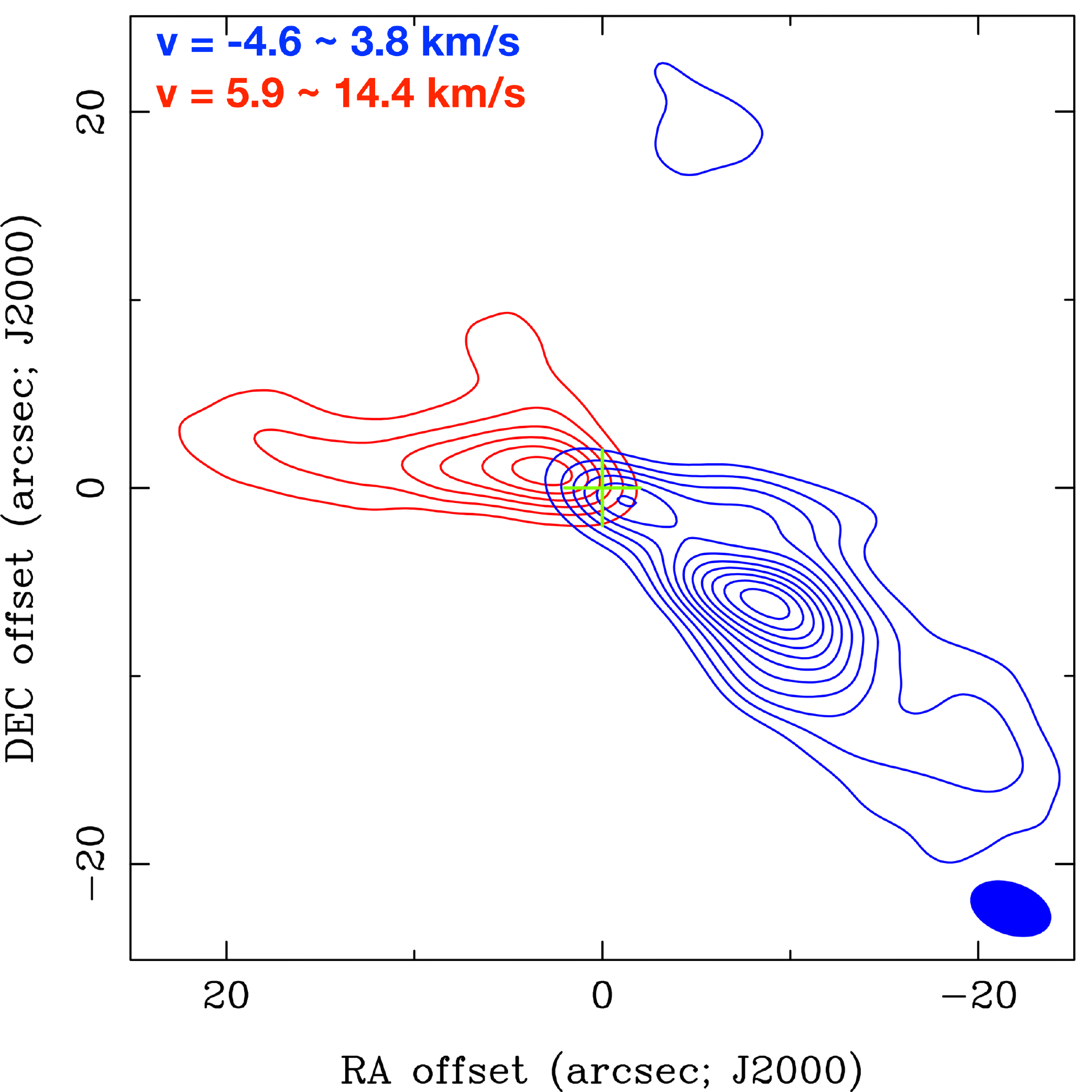}
\caption{$^{12}$CO (2-1) moment 0 map with blue contours integrated from $v = −4.6$ km s$^{-1}$ to $v = 3.8$ km s$^{-1}$ and red contours from $v = 5.9$ km s$^{-1}$ to $v = 14.4$ km s$^{-1}$. A filled ellipse in the bottom right corner denotes the synthesized beam size. A cross shows the protostellar position. Contour levels are from 3$\sigma$ in steps of 3$\sigma$, where 1$\sigma$ is 1 Jy beam$^{-1}$ km s$^{-1}$.}
\label{fig:12}
\end{figure}

Figure \ref{fig:12} displays the SMA $^{12}$CO (2--1) line emission in L1455 IRS1. 
A V-shaped structure is revealed in the red-shifted wing with an opening angle of $\sim$45$^\circ$. The blue-shifted wing is also converging to the protostellar position, with a less pronounced geometry but a likely opening angle similar to its red-shifted counterpart. The blue-shifted wing extends to the southwest over $\sim 25''$ in length ($\sim$6250 AU). 
The shorter ($\sim 10''$) red-shifted wing is pointing to the northeast. 

The $^{12}$CO emission likely outlines the collimated bipolar outflow as observed previously in H$_2$ \citep{davis19972} and $^{12}$CO (3--2) with the JCMT \citep{curtis20102}. Their single-dish maps suggest a $\sim$40$^\circ$ P.A. of the outflow axis, while the recent interferometric observation by \citet{hull2014} with CARMA shows a slightly larger P.A. of $\sim$66$^\circ$
(measured counter-clockwise from north). Our $^{12}$CO emission map with the SMA aligns closely with the CARMA result. We, therefore, adopt 66$^\circ$ as the outflow axis P.A. in the following analyses. 

\subsubsection{C$^{18}$O (2--1) Emission}\label{sec:18sma}
Figure \ref{fig:m0}(c) shows the integrated-intensity (i.e., moment 0) map (contours) overlaid on the intensity-weighted mean velocity (i.e., moment 1) map (color)
of the C$^{18}$O (2--1) emission in L1455 IRS1.
Generally, the emission delineates a compact morphology with a size of $\sim$1000 AU. The emission above 6$\sigma$ shows a clear velocity gradient perpendicular to the $^{12}$CO (2--1) outflow axis (Figure \ref{fig:12}).
A two-dimensional Gaussian is fitted to the central compact region with intensity $>5\sigma$ giving a deconvolved size of $\sim6''$. 
Furthermore, another velocity gradient is present in the northwestern part, with a gradient opposite to the one across the innermost region. 
A protrusion extending toward the southwest is lying approximately perpendicularly to the larger-scale velocity gradient. Given its direction that is roughly aligned with the outflow axis, this could be an outflow contamination.
A second smaller protrusion is found to the north. 

\begin{figure*}[!ht]
\centering
\includegraphics[width=\textwidth]{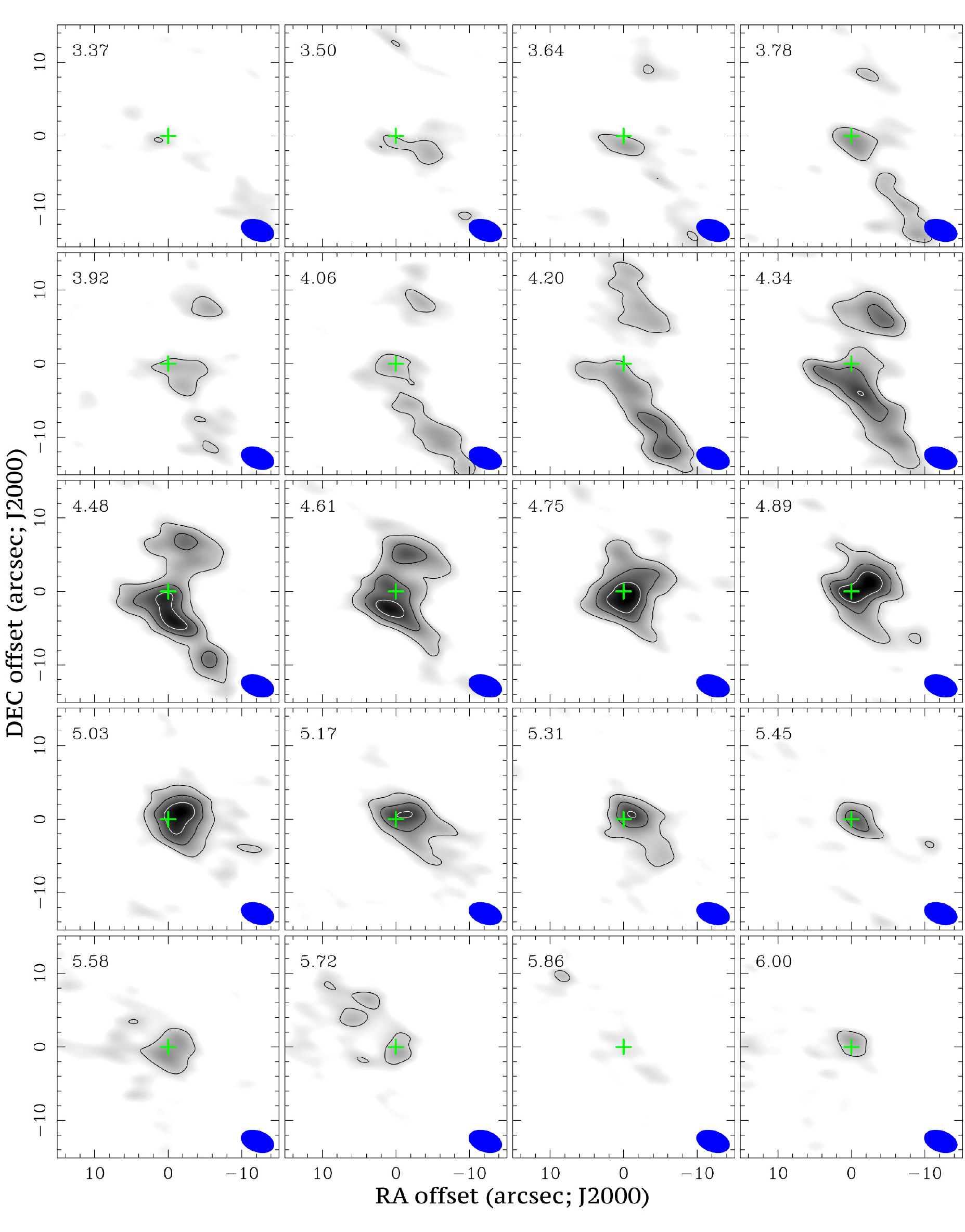}
\caption{SMA velocity channel maps of C$^{18}$O (2--1) emission in L1455 IRS1 (two channels binned). The velocity of each channel is shown in the upper left corner in each panel. A filled ellipse in the bottom right corner in each panel denotes the beam size. Crosses show the IRS1 protostellar position. Contour levels are from 3$\sigma$ in steps of 3$\sigma$, where 1$\sigma$ is 115 mJy beam$^{-1}$.}
\label{fig:cbch}
\end{figure*}

Figure \ref{fig:cbch} shows the SMA velocity channel maps of C$^{18}$O (2--1) centered on L1455 IRS1. 
We bin two channels for higher signal-to-noise ratio.
20 channels have a maximum emission above a 3$\sigma$ level, with velocities ranging from $3.26$ km s$^{-1}$ to $6.13$ km s$^{-1}$. At velocities $\lesssim 4.8$ km s$^{-1}$, the emission indicates a compact object located south to the protostellar position. In the range $4.8 \lesssim v \lesssim 5.3$  km s$^{-1}$, this emission is shifted north. Around the systemic velocity of about 4.7 km s$^{-1}$, 
the emission stretches to the north and the southwest, along the outflow axis.
For velocities larger than $\sim$5.3 km s$^{-1}$, however, the emission shows no significant spatial offsets. This demonstrates an asymmetrical velocity distribution along the possible rotational axis, likely due to the two protrusions joining the compact object from the north and the west. 

\subsection{IRAM - C$^{18}$O (2--1) Emission} \label{sec:iram_res}
\begin{figure*}[!ht]
\centering
\includegraphics[width=\textwidth]{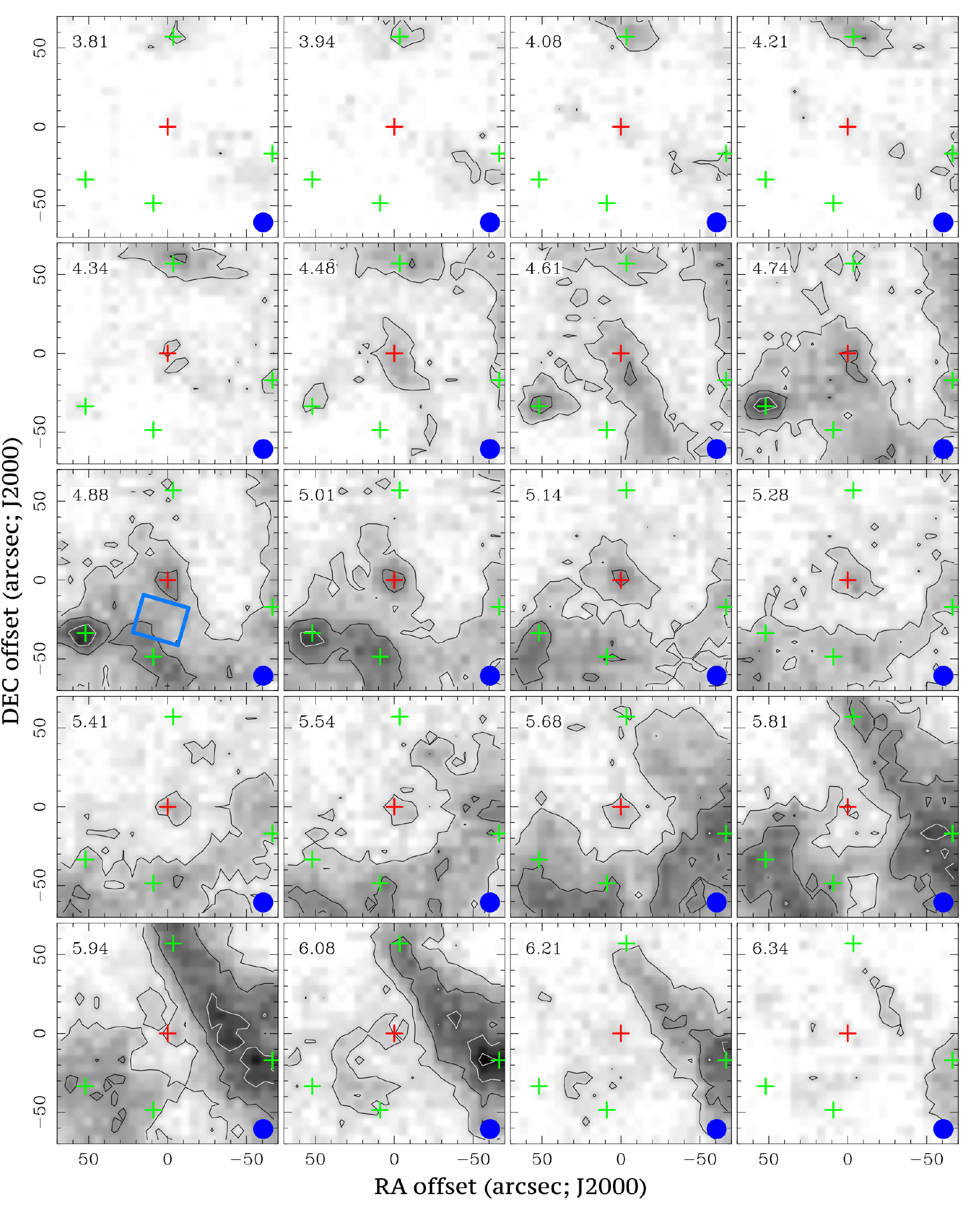}
\caption{IRAM-30m C$^{18}$O (2--1) velocity channel maps around L1455 IRS1 (five channels binned). The central velocity of the five binned channels is shown in the upper left corner in each panel. A filled circle in the bottom right corner in each panel denotes the beam size. Red crosses show the protostellar position of L1455 IRS1, green ones indicate the positions of IRS4, IRS5, HRF40, and CoreW. A blue box in the middle left panel indicates the location of the filamentary structure in between IRS1 and HRF40. Contour levels are from 3$\sigma$ in steps of 3$\sigma$, where 1$\sigma$ is 0.107 K.}
\label{fig:iram}
\end{figure*}

Figure \ref{fig:m0}(a) displays moment 0 and moment 1 C$^{18}$O (2--1) overlaid maps observed with the IRAM-30m telescope. The map includes the central protostar L1455 IRS1, and partly covers two more protostars, L1455 IRS4 and IRS5, and two possible starless cores, HRF40 \citep{hatchell20071,curtis20101} and CoreW (this paper). 
We present C$^{18}$O (2--1) velocity channel maps (5 channels are binned for presentation) in Figure \ref{fig:iram}. In the regions between the cores, significant emission is detected that
appears to form filamentary structures and bridges connecting the individual cores.
The central component observed in the moment 0 map is best described by a core size of $\sim40''$ from two-dimensional Gaussian fitting.

Multiple velocity components can be seen at and connecting to the locations of L1455 IRS4 and IRS5. 
At the location of IRS4, a first component around $v\sim 4.8$ km s$^{-1}$  is likely a protostellar core, whereas a second component around $v\sim$5.8 km s$^{-1}$ might correspond to a larger structure which extends to the southwest.
Indeed, most studies in the literature measure a systemic velocity for L1455 IRS4 of $\sim$4.9 km s$^{-1}$, and not 5.8 km s$^{-1}$ \citep{juan1993,kirk2007,friesen2013}. Towards L1455 IRS5, a bridge protruding from CoreW can be readily seen from $v\sim5.5$ km s$^{-1}$ to $\sim6.2$ km s$^{-1}$. From JCMT $^{13}$CO observations \citep{curtis20101}, this bridge appears to be linked to IRS5 because it doest not stretch further across its position. However, a second more compact and uniform (in shape) component with a centroid velocity of $\sim$4.4 km s$^{-1}$ lies on the IRS5 protostellar position as well. 

L1455 IRS1 appears to be marginally elongated along a north-south axis.
A unique velocity gradient linking IRS1 and HRF40, with a P.A. $\sim$164$^\circ$, occurs from $v\sim4.8$ km s$^{-1}$ (IRS1) to $\sim$5.0 km s$^{-1}$ (Section \ref{sec:conn_fila}). In the IRAM-30m C$^{18}$O emission, this is the only manifest large-scale structure that is connected with IRS1 in both space and velocity. Although clear in velocity, it is mixed spatially with surrounding structures, particularly the
triangular zone around IRS1. The SMA C$^{18}$O emission shows no clear corresponding structure, implying that this large-scale structure is either resolved out by the SMA or it is not associated with the inner structures directly. 

\subsection{SMA-IRAM-Combined C$^{18}$O (2--1) Emission}\label{sec:ressec}
\begin{figure*}[!ht]
\centering
\includegraphics[width=0.8\textwidth]{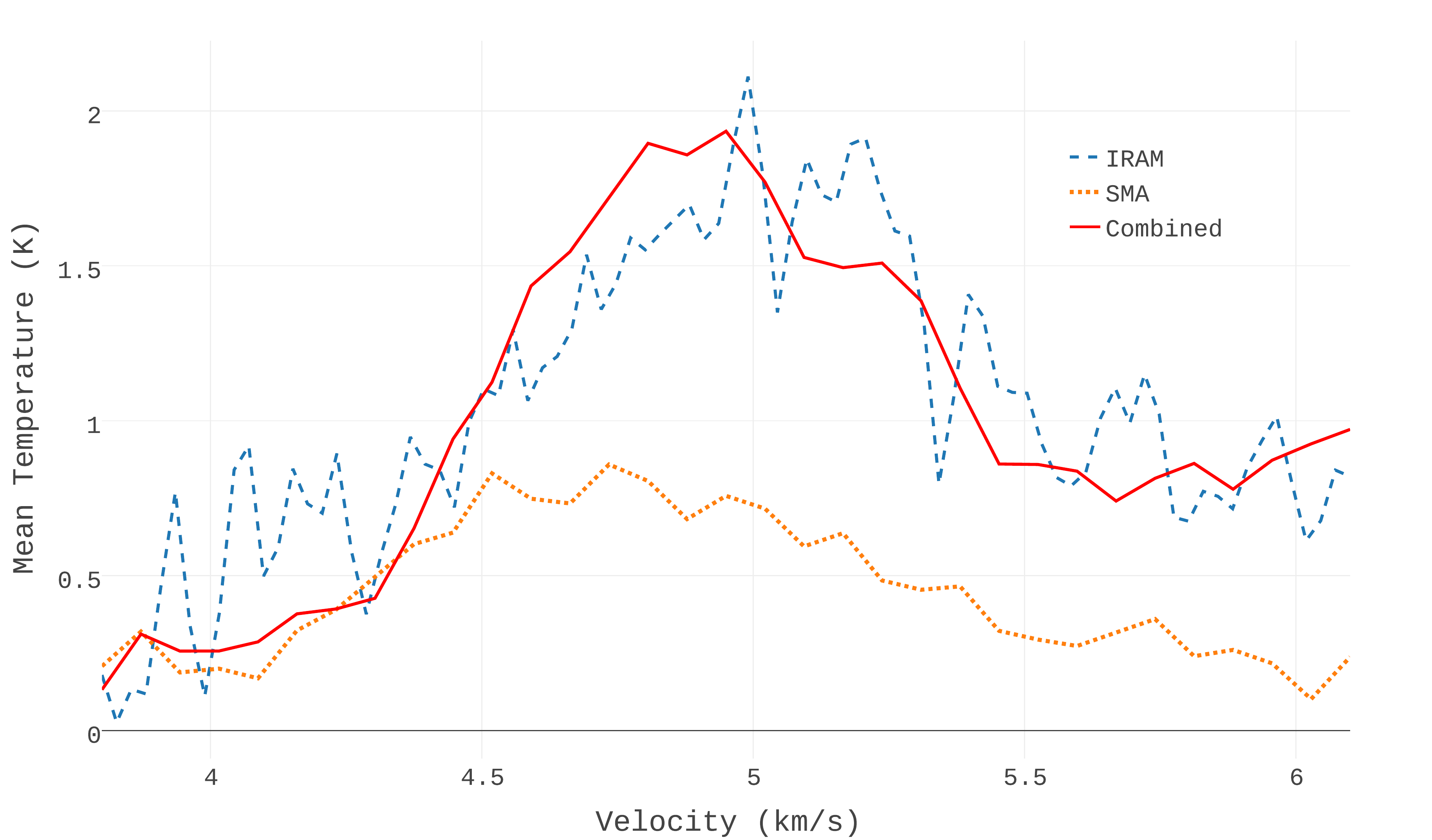}
\caption{C$^{18}$O (2--1) spectra in the central $5\farcs6 \times 5\farcs6$ (IRAM central pixel) of L1455 IRS1 with respect to velocity. The solid red curve shows the spectrum of the SMA-IRAM-combined map. The blue dashed and orange dotted curves are the IRAM- and SMA-only spectra.}
\label{fig:cbfx}
\end{figure*}

Figure \ref{fig:cbfx} shows the flux comparison for the C$^{18}$O (2--1) emission in the central $5\farcs6\times5\farcs6$ region from the SMA, IRAM-30m, and the SMA-IRAM-combined maps. The SMA has about 70\% missing flux in velocities $\gtrsim$4.8 km s$^{-1}$. The combined map recovers roughly all the flux as observed by the IRAM-30m. 

The SMA-IRAM-combined moment 0 and moment 1 maps of C$^{18}$O (2--1) are overlaid in Figure \ref{fig:m0}(b). In the innermost region, the velocity gradient perpendicular to the outflow direction seen in the SMA-only map (Figure \ref{fig:m0}(c)) is still present but less distinct.
The large gradient seen in the IRAM-only map extending to the south (from IRS1 to HRF40, Section \ref{sec:conn_fila}) can be seen in the combined map as the blue-green region connecting towards the central SMA component.
While both gradients are observed in the combined map, it is obvious that the combination process has added more detailed information that, in this case, tends to blur features that are clearly detected in the SMA- and IRAM-only maps.
This is explained by noticing that the IRS1 disk and the filament 
are well captured by the separate scales probed by the SMA and IRAM-30m, respectively. This is not obvious from the beginning and will generally depend on a source size and its distance. 
Additionally, the combined map demonstrates that there are only smaller-scale but no coherent larger-scale structures that could bias identification and interpretation of the large filament connecting towards IRS1 (Section \ref{sec:conn_fila}).

\section{Analysis}\label{sec:ana}
\subsection{Position-Velocity Diagrams: Rotation in L1455 IRS1}\label{sec:pvdiagram}

\begin{figure*}[!ht]
\centering
\includegraphics[width=0.99\textwidth]{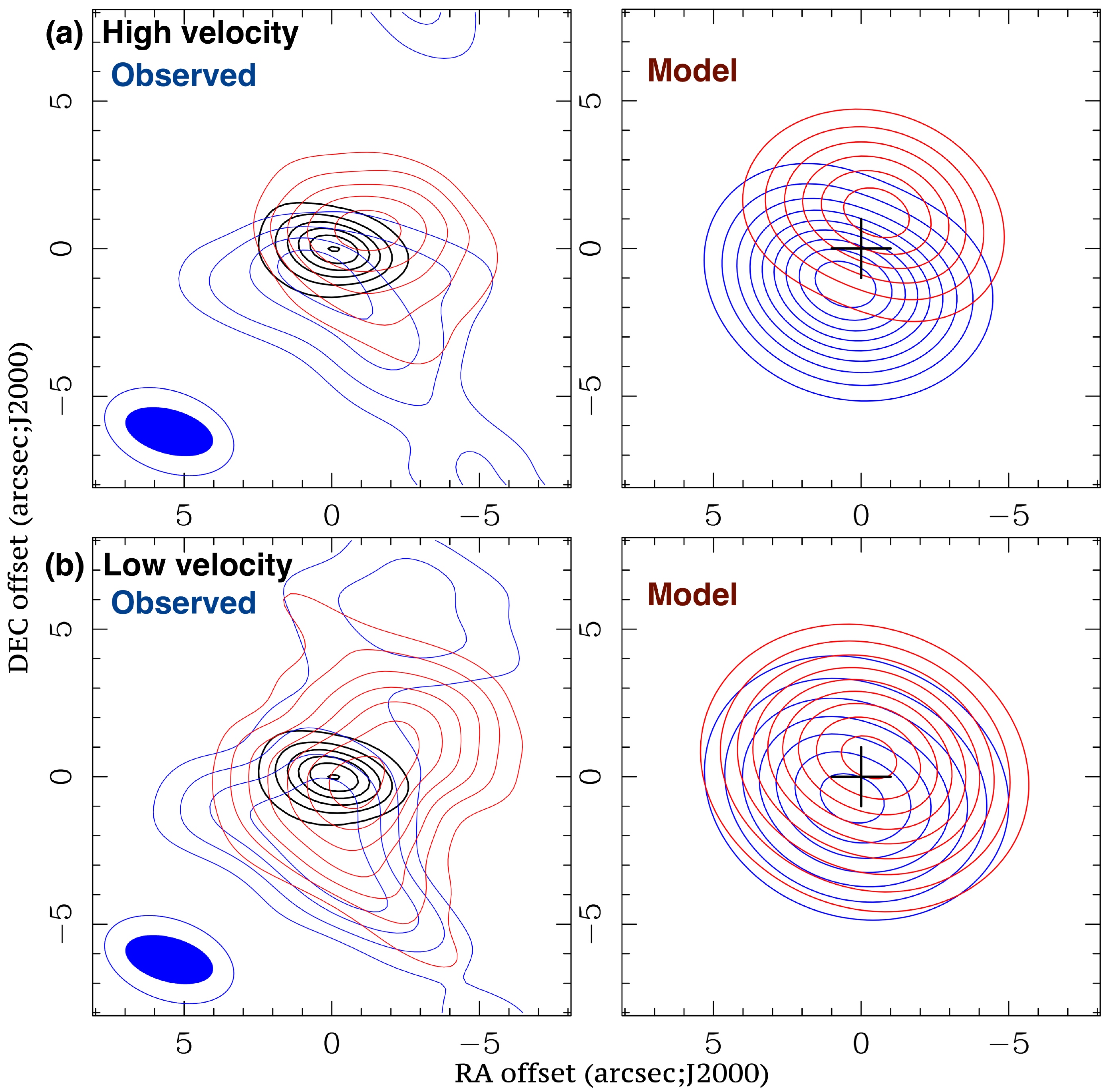}
\caption{\textit{Left panels:} Comparison between the SMA 1.3 mm continuum (black) and the SMA blue- and red-shifted C$^{18}$O (2-1) emission. \textit{Right panels:} Corresponding images of the Keplerian model. (a) High-velocity emission, (b) low-velocity emission. The velocity ranges (relative to $V_\text{sys}$) are 0.7--1.0 km s$^{-1}$ and $<0.7$ km s$^{-1}$ for high- and low-velocity, respectively. The continuum beam is shown at the bottom left corner. The larger ellipse shows the beam size for the C$^{18}$O emission. Crosses show the protostellar position. Contour levels are from 5$\sigma$ in steps of 5$\sigma$ for continuum, 2$\sigma$ for C$^{18}$O (2-1). }
\label{fig:rot}
\end{figure*}

YSOs evolve via accreting material. The usually accompanying outflows are believed to be the removal mechanism for excessive angular momentum (e.g., \citet{MO2008}). 
Due to the distinct outflows in L1455 IRS1, detected in $^{12}$CO (2--1) (Figure \ref{fig:12}), we expect that the core is also associated with a disk that we assume to be rotating in a plane orthogonal to the outflow axis.
The optical depth of the C$^{18}$O (2-1) emission in L1455 IRS1 is estimated to be $\sim0.2$ on average in a velocity channel in the central 2$''\times2''$ region of the combined map,
where the excitation temperature is taken to be 12.6K \citep{curtis20101}.
The emission is, thus, optically thin and tracing the innermost parts in the envelope.
In conclusion, the C$^{18}$O (2-1) emission is indicative of the motions near the protostar. 

\begin{figure*}[!ht]
\centering
\includegraphics[width=\textwidth]{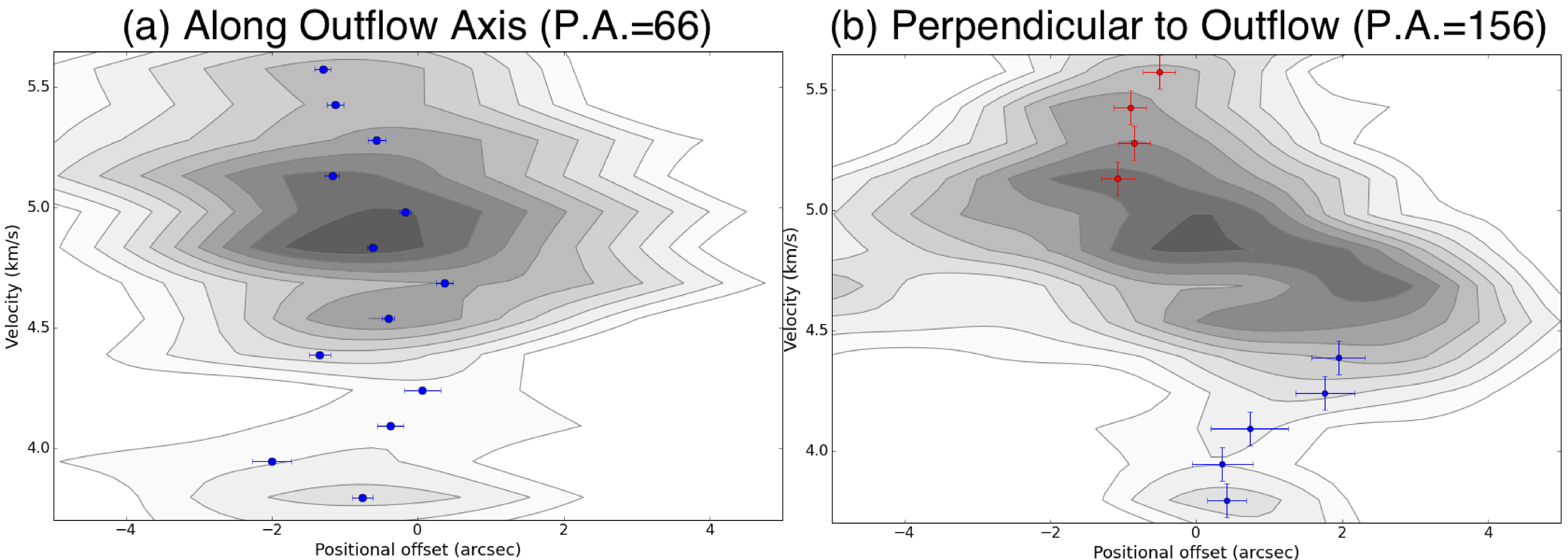}
\caption{Position-Velocity diagrams of the C$^{18}$O (2--1) SMA map of L1455 IRS1, (a) along and (b) perpendicular to the outflow axis.
The absence of velocity gradients in (a) implies the absence of non-rotational motions along the radial direction of the disk.
The blue and red points in (b) stand for blue- and red-shifted velocities with respect
to the systemic velocity. 
Contour levels are from 3$\sigma$ in steps of 1$\sigma$, where 1$\sigma$ is 115 mJy beam$^{-1}$.}
\label{fig:pv}
\end{figure*}

The left panel in Figure \ref{fig:rot}(a) presents a comparison between the 1.3 mm continuum and the high-velocity C$^{18}$O (2--1) emission. 
The blue- and red-shifted high-velocity components and the protostar are well aligned along the axis perpendicular to the outflow direction, hinting the existence of a rotational motion.
Figure \ref{fig:pv} displays Position-Velocity (P-V) diagrams of the C$^{18}$O (2-1) emission of the SMA map of L1455 IRS1. The center of the P-V diagrams is the protostellar position. The P-V cuts are along P.A.$=66^\circ$ in Figure \ref{fig:pv}(a), and P.A.$=156^\circ$ in Figure \ref{fig:pv}(b), which are the angles along and perpendicular to the outflow axis (see section \ref{sec:12co}). 

Since the line-of-sight velocity of a purely rotating disk is zero along the minor axis, any velocity variation along this axis must be non-rotational. In other words, the velocity profile along the minor axis can be a measure for infalling motions or outflow contamination.
If there is infalling motion, we expect a velocity gradient along the minor axis. Therefore, the lack of a significant velocity gradient in Figure \ref{fig:pv}(a) suggests that there is no clear infalling motion in IRS1.
On the other hand, for any disk that is not face-on, i.e., with an inclination angle different
from $90^\circ$, the velocity profile along the major axis represents rotational disk features. Thus, we are 
aiming at measuring velocity profiles as a function of rotational radius along each axis. 
Figure \ref{fig:pv}(b) infers a differential velocity increase from larger to smaller rotational radii. Combining these findings, we conclude on the existence of a dominant rotating structure in this region (e.g., \citet{belloche2013}). 
We quantify this rotation in the subsequent section, following the technique outlined in \citet{yen2013}.

We note that because of asymmetric envelope features (see Section \ref{sec:nds} below), it is difficult to reproduce all the observed patterns with a simple model. 
As we will show in section \ref{sec:nds}, these structures cannot be described by a flat disk model.
In the P-V diagrams described here, most of these structures either lie outside of the diagram (i.e., located at $>5''$ from the center), or they are much weaker in terms of emission intensity as compared to the rotational feature. 

\subsection{Existence of Disk}\label{sec:disk}
We assume a velocity profile for rotation that follows a power law
\begin{align}\label{eqn:pl}
v_\text{rot}(r) \propto r^p,
\end{align}
where $v_\text{rot}$ is the rotational velocity at a radius $r$, and $p$ is the power law index. Projected on the plane-of-sky, $p$ does not change, but the rotational velocity and radius are replaced by the projected relative velocity ($\equiv V$) and positional offset in the P-V diagram. We, thus, fit the representative points in Figure \ref{fig:pv}(a) with Equation (\ref{eqn:pl}) to derive $p$ and measure the disk rotation.

\begin{figure}[ht!]
\centering
\includegraphics[width=0.49\textwidth]{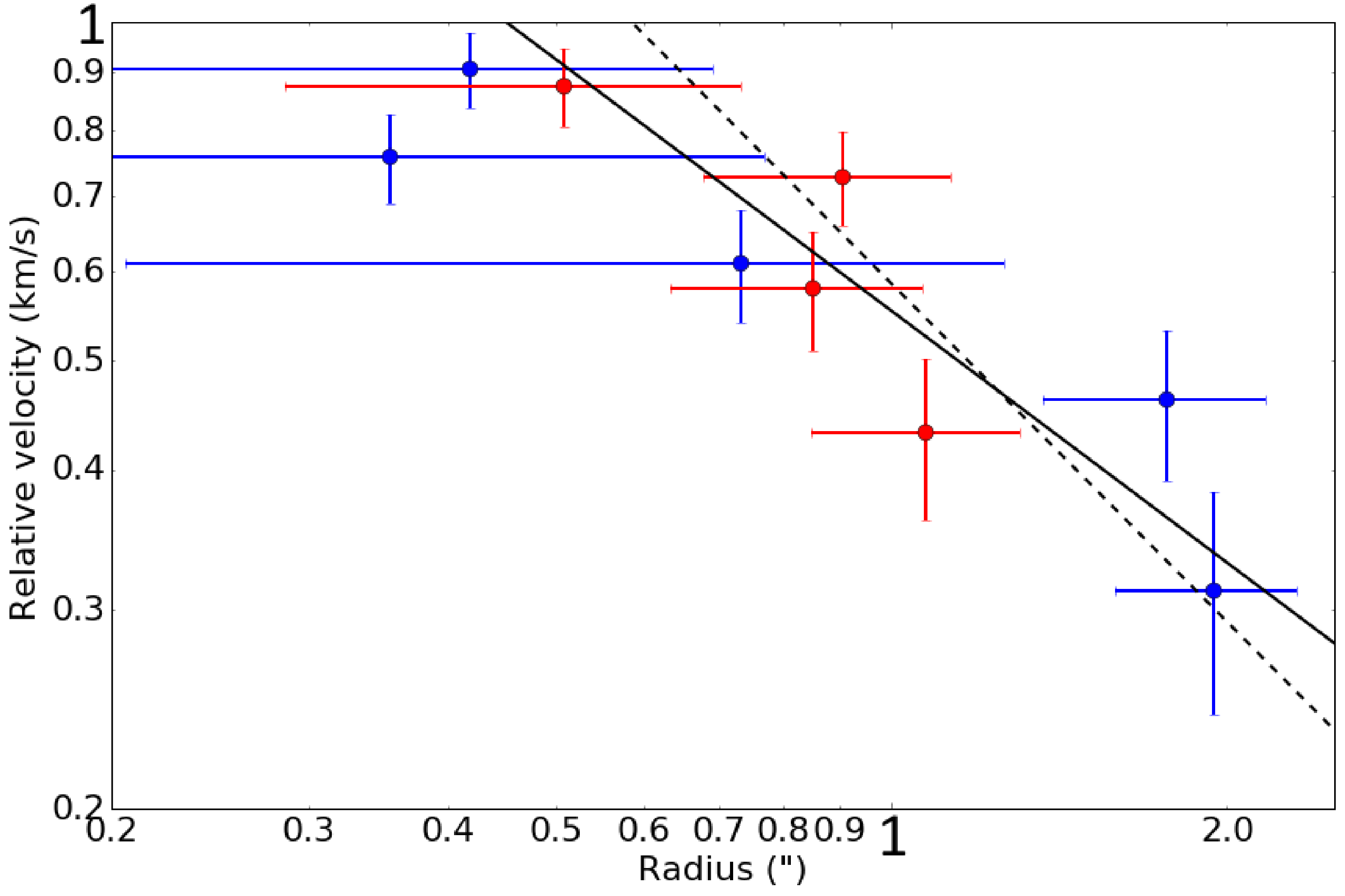}
\caption{Projected velocity profile of L1455 IRS1 as a function of radius $r$ in log-log scale. Blue and red points stand for blue- and red-shifted velocities with respect to the systemic velocity. The best-fit power law index is $p=-0.75 \pm 0.16$ with a systemic velocity $V_{sys}=4.70$ km s$^{-1}$ (black solid line). 
The black dashed line indicates the best-fit result for a fixed value $p=-1$ for the outer 6 points.}
\label{fig:pl}
\end{figure}

The systemic velocity is found by fitting Gaussian profiles to the spectra used for Figure \ref{fig:cbfx}. 
The SMA and IRAM-30m spectra give 4.70 km s$^{-1}$ and 4.99 km s$^{-1}$, respectively. 
This implies that the two maps are tracing different gas components.
This further explains why rotation is not evident in the combined map in Figure \ref{fig:m0}(b).
The larger scale components with higher velocities in the IRAM-30m map
are likely structures that are not directly associated with the inner rotating SMA component.
Consequently, components from outer regions cannot be used to analyzing rotation.
We, therefore, do not include the IRAM-30m map for more outer data points in the P-V diagrams, and
we consider only the SMA map be tracing the rotational motion in the inner region.
Hence, the systemic velocity ($\equiv V_\text{sys}$) is fixed to 4.70 km s$^{-1}$.
Error bars in velocities are set to be the channel width. 

To determine the profile, we fit a Gaussian in each velocity channel to get a centroid position which is then regarded as the rotational radius at that rotational velocity.
A two-dimensional orthogonal distance regression is applied to find the best-fit parameters as shown in Figure \ref{fig:pl}.
As a result, the observed motion can be best represented by a power law with $p=-0.75 \pm 0.16$.
This suggests that the rotational features seen in the P-V diagram are likely caused by a spin-up rotation.

Observations show that protostellar disks can possess two distinct motions (e.g., \citet{murillo2013,yen2015}): infalling (and rotating) motion with conserved angular momentum ($p=-1$), and pure Keplerian rotation ($p=-0.5$).
A power law index in between these two values is possibly caused by insufficient instrumental resolution,
blending the two types of motion and rendering them indistinguishable (e.g., \citet{lee2010}). 
Indeed, flattened inner data points can be seen in the velociy profile. 
This is also observed in L1527 IRS, and possibly indicates the existence of such a transition \citep{ohashi2014}. 
To provide further evidence,
we fit another power law only to the outer 6 points, obtaining an index of $-0.9 \pm 0.3$,
which is suggestive of an infalling (and rotating) envelope with conserved angular momentum.
Thus, assuming the rotation indeed has a constant angular momentum, 
we force $p=-1.0$ and fit for the outermost 6 data points.
The result is shown in Figure \ref{fig:pl},
likely hinting the presence of a Keplerian disk with a radius $\lesssim 0\farcs8$ ($=$200AU) in IRS1. We note that this radius is consistent with the deconvolved size of the SMA 1.3mm continuum image (Section \ref{sec:con}), and hence we adopt a disk radius of 0\farcs8.
As will be discussed below, the inclination angle of L1455 IRS1 is likely in the range of $22.5^\circ$ to $67.5^\circ$. Assuming it to be  $45^\circ$,
the best-fit profile gives a specific angular momentum $j=(1.0\pm0.09) \times 10^{-3}$ km s$^{-1}$ pc.
Moreover, for the inner disk radius of $0\farcs8$, the protostellar mass ($\equiv M_\ast$) can be calculated following Keplerian motion, $v_\text{rot}=\sqrt{GM_\ast/r}$,
which gives $M_\ast = (0.28\pm0.05) M_\sun$.

\subsection{Asymmetric Envelope Features}\label{sec:nds}
\begin{figure}[!ht]
\centering
\includegraphics[width=0.49\textwidth]{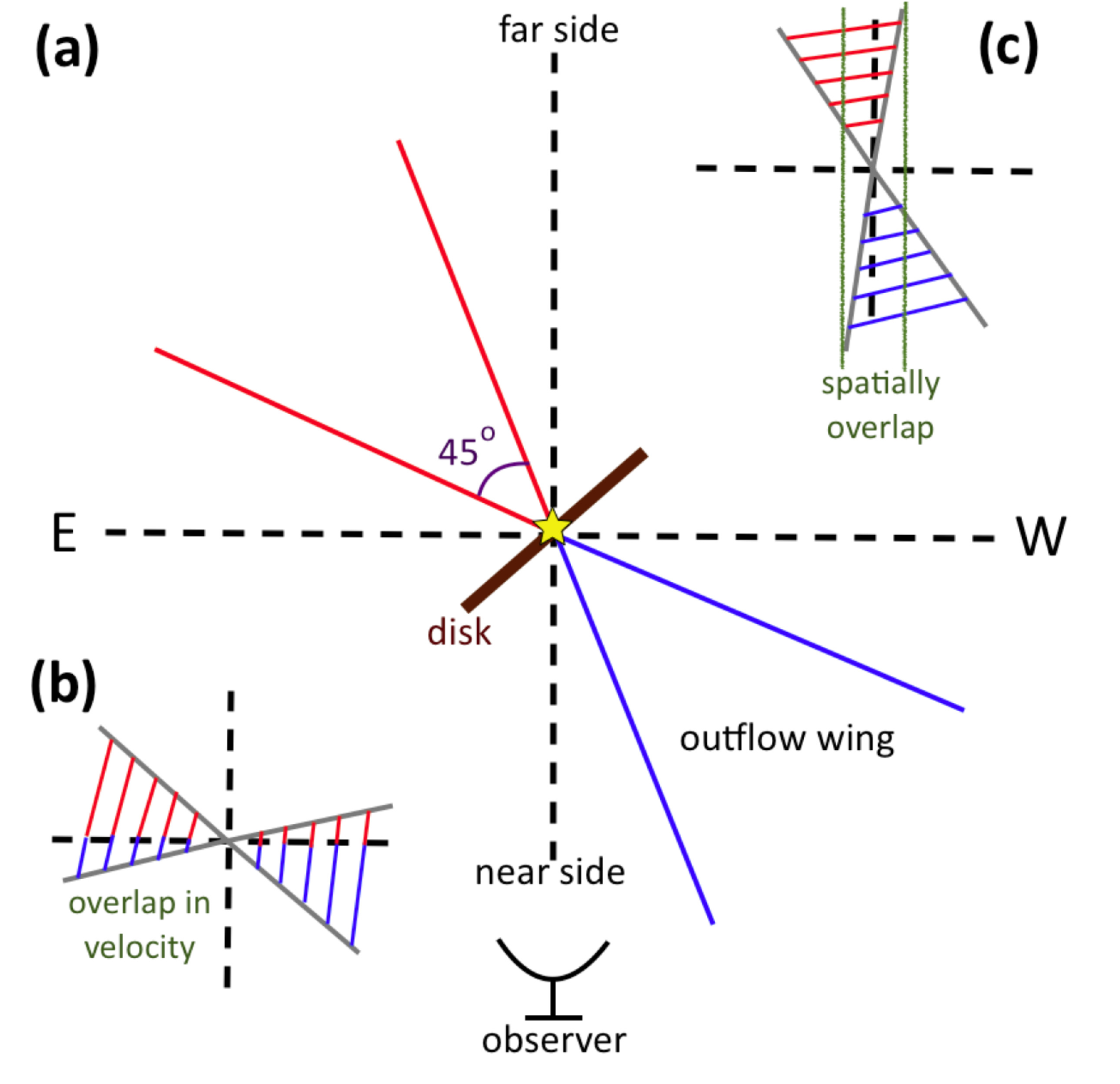}
\caption{Schematics of outflow wings. The opening angle projected on the plane is $\sim 45^\circ$. (a) The observed outflows show two distinctive separate wings, red- and blue-shifted. (b) If $i$ is larger than $67.5^\circ$, each wing shows both red- and blue-shifted emission. (c) The wings near the central part on the plane-of-sky overlap. They, thus, appear as overlapping red- and blue-shifted emission in the central region, if $i$ is less than $22.5^\circ$.}
\label{fig:outgeo}
\end{figure}

In section \ref{sec:disk}, we conclude that the rotation in L1455 IRS1 is hinting a disk that is likely of Keplerian nature in its inner part. 
In this section, we further examine the properties of the envelope of L1455 IRS1. 
We construct a simple Keplerian disk model, which is parametrized by four physical quantities: 
inclination angle ($\equiv i$), position angle of the major axis ($\equiv \theta$), 
two-dimensional intensity distribution on the plane-of-sky, and rotational velocity distribution. 
Since the red- and blue-shifted wings of the $^{12}$CO (2-1) outflow do not overlap spatially except for the central compact structure, 
the permitted range of inclination angles can be inferred based on geometrical arguments. 
Figure \ref{fig:outgeo} illustrates the ideas. Suppose the outflow wings are perfect cones. 
Hence, the opening angle projected on any plane is the same as that on the plane-of-sky. 
From the geometry of the integrated intensity map, the opening angle is estimated to be about $45^\circ$.
By noticing that the outflow axis is orthogonal to the disk major axis, 
the inclination angle can be constrained from the distribution of red- and blue-shifted emission on the plane-of-sky. 
The wings start to overlap in velocity (Figure \ref{fig:outgeo}(b)) if the disk is nearly edge-on, i.e., 
$i \geqslant 90^\circ-45^\circ/2=67.5^\circ$. They are spatially overlapped along the line-of-sight (Figure \ref{fig:outgeo}(c)) 
if $i \leqslant 45^\circ/2=22.5^\circ$.
Otherwise the two wings are completely separated, both in terms of spatial distribution and velocity \citep{ulrich1976}.

We set $i=45^\circ$, because the geometrical appearance of the C$^{18}$O (2-1) moment 0 map (Figure \ref{fig:m0}(b)) 
is neither round nor thin, and the red- and blue-shifted emission barely overlap in the central part. 
Specific angular momentum and protostellar mass are then $j=(1.0 \pm 0.09)\times 10^{-3}$ km s$^{-1}$ pc
and $M_\ast = 0.28 \pm 0.05 M_\sun$ (Section \ref{sec:disk}).
The slight overlapping of the red- and blue-shifted wings is likely due to insufficient spatial resolution, as a result of beam convolution.
The position angle is fixed to $156^\circ$, as also used for the P-V diagrams.
The deconvolved size of the two-dimensional Gaussian of the central part of the SMA C$^{18}$O (2-1) integrated intensity map 
(Figure \ref{fig:m0}(c), Section \ref{sec:18sma}) is adopted as the intensity distribution for the disk model. 
By doing so, we assume that the disk is completely axisymmetric and most of the emission is dominated by the disk. 
With this, a simple Keplerian model is constructed with the MIRIAD tasks \textit{velmodel} and \textit{velimage}, 
where the rotational velocity profile is determined by 
the above $M_\ast$. 

The results are illustrated in Figure \ref{fig:rot}.
A comparison between observations and the model for high-velocity emission (Figure \ref{fig:rot}(a)) shows an overall consistency: 
The blue- and red-shifted high-velocity components are likely tracing the rotation of the inner Keplerian disk. On the other hand,
the low-velocity channels (Figure \ref{fig:rot}(b))
present the asymmetric envelope:
In both red- and blue-shifted components, 
an elongated structure lying approximately along the outflow direction can be seen. 
This feature is likely caused by outflow contamination. 
A northern clump extending along the disk major axis to $\sim6''$ away from the protostellar position is evident in the blue-shifted channels. 
As mentioned in section \ref{sec:18sma}, this feature shows a velocity gradient inverse to the one across the disk.
It, therefore, likely is independent of the rotating disk. 
These structures are, thus, features in the surrounding envelope, 
which cannot be explained by a simple Keplerian rotation.

\section{Discussion}\label{sec:dis}

\subsection{Fast Rotation in Small Core}\label{sec:fast}

\begin{figure}[ht!]
\centering
\includegraphics[width=0.49\textwidth]{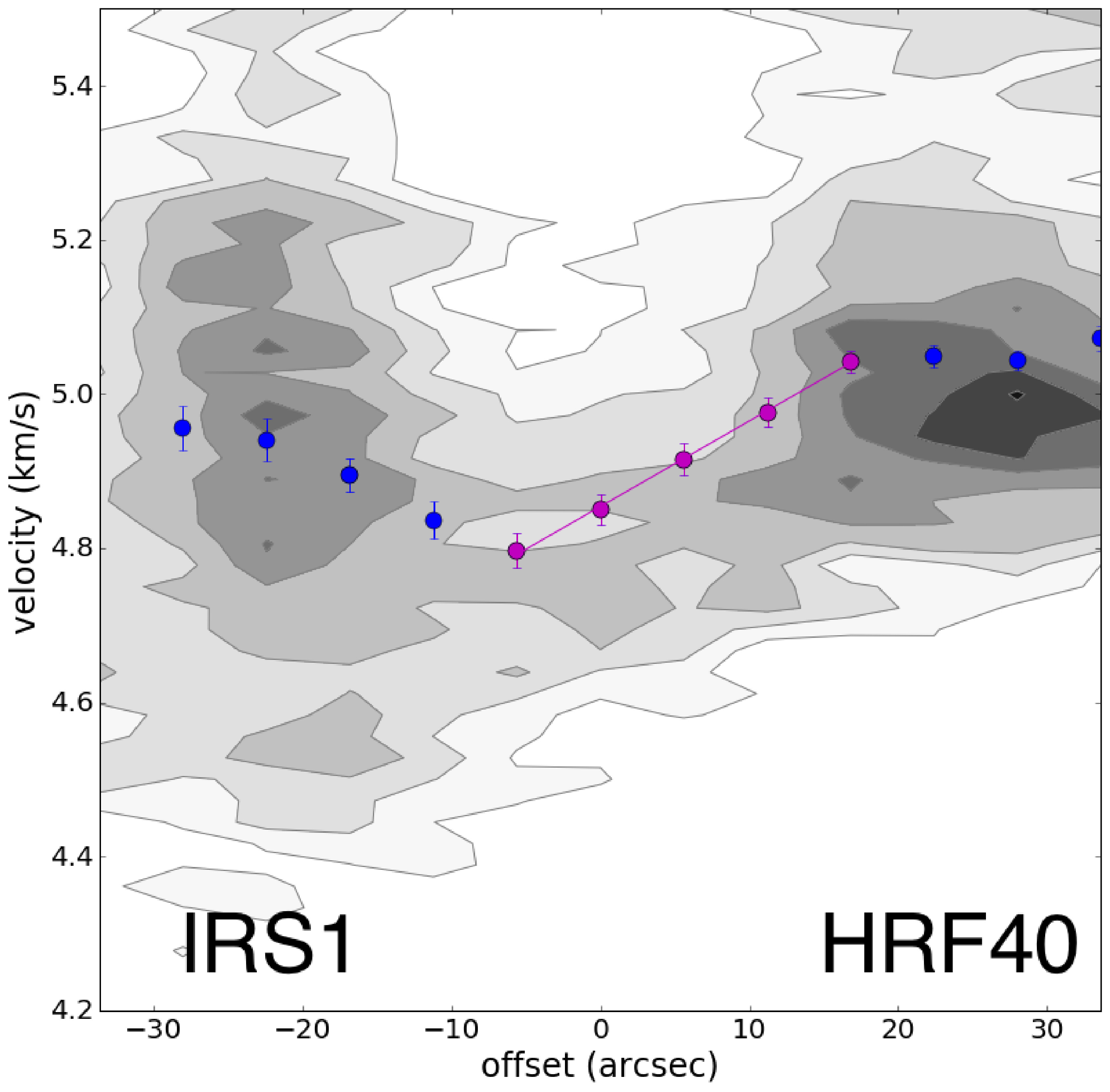}
\caption{PV diagram of the IRAM C$^{18}$O (2--1) emission along P.A.$=164^\circ$. The protostar L1455 IRS1 lies on the left side, whereas the peak on the right side is the starless core HRF40. A filamentary structure linking the two cores is marked by the magenta points, which indicate the velocity gradient along the filament. The velocity gradient is 8.1$\pm$0.8 km s$^{-1}$ pc$^{-1}$. The contours are from 3$\sigma$ to 9$\sigma$ in step of 1$\sigma$, where 1$\sigma$ is 0.156K.}
\label{fig:filagd}
\end{figure}

To quantify rotation in the IRS1 core,
we measure the centroid velocities of the core along P.A.$=156^\circ$ 
(i.e., along the disk major axis), by fitting Gaussian functions 
to the spectra at radii from 0 to $\sim2000$AU in the IRAM-30m data.
A linear velocity gradient is fitted to the profile, yielding (7.8$\pm$5.2) km s$^{-1}$ pc$^{-1}$.
This gradient suggests a rotation with the same direction as the disk and envelope (Section \ref{sec:ana}).
Assuming the core follows a rigid body rotation, 
the specific angular momentum $j$ amounts to $(4.6\pm3.1)\times 10^{-3}$ km s$^{-1}$ pc, 
at the radius of the core (i.e., 5000AU, see Section \ref{sec:iram_res}).
\citet{goodman1993} find a power-law relation between core sizes and specific angular momenta 
in complex dense cores, following 
$j = 10^{-0.7\pm0.2}R^{1.6\pm0.2}$, where $j$ and $R$ are in units of km s$^{-1}$ pc and pc, respectively.
Specifically, for a core similar in size to IRS1 ($R=5000$AU$\sim0.024$pc), the relation predicts $j = (5.2\pm4.5)\times 10^{-4}$ km s$^{-1}$.
\citet{tobin2011} have observed velocity gradients of $\sim1$ km s$^{-1}$ pc$^{-1}$, perpendicular to outflow axes in typical Class 0 cores.
These measured gradients indicate $j \sim6\times10^{-4}$ km s$^{-1}$ pc on a 5000AU scale,
consistent with the relation of \citet{goodman1993}.
Hence, our observed value is one order of magnitude larger than expected.
This suggests that the core possesses an uncommonly large specific angular momentum for its size.
As we discuss in the following section, the surrounding filamentary structure might be responsible for this larger than average momentum.

\subsection{Connection to Large-Scale Filament}\label{sec:conn_fila}

\begin{figure*}[!ht]
\centering
\includegraphics[width=.99\textwidth]{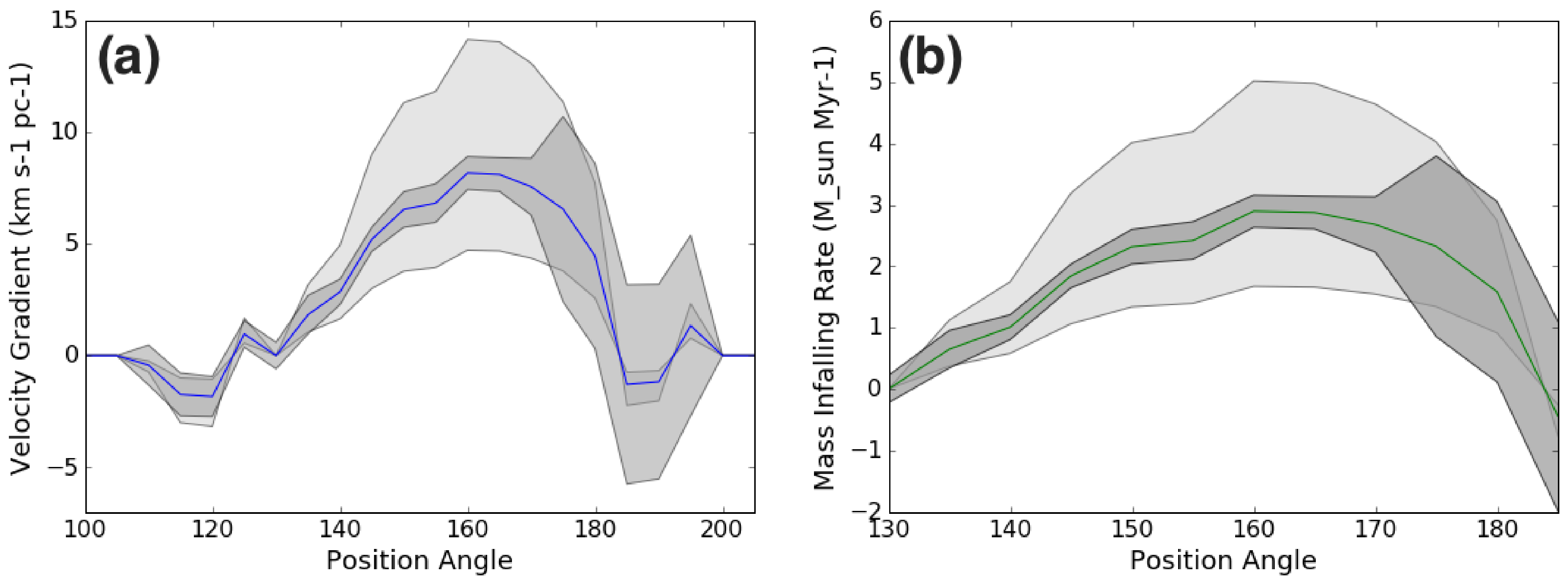}
\caption{(a) Measured velocity gradient as a function of position angle extending from IRS1. (b) Mass infalling rate along each position angle of the filament. Colored lines indicate the best-fit gradient and the corresponding mass infalling rate. Dark gray shaded areas show error bounds from the fitting. A projection effect of $45^\circ\pm 15^\circ$ is represented by the light gray area. }
\label{fig:fila_grad}
\end{figure*}

Figure \ref{fig:filagd} is the P-V diagram along the structure extending from L1455 IRS1 (Figure \ref{fig:iram}). This structure shows emission above $5 \sigma$, with a linear gradient in the region between the two protostellar cores IRS1 and HRF40. 
We fit for centroid velocities at each positional offset to get a velocity profile. 
The extracted profile either changes sign in its slope (towards IRS1) or it flattens (towards HRF40).
We interpret the linear section of the gradient as a filament of length $\sim 0.034$pc (e.g., \citet{hacar2011,kirk2013}). Best-fit magnitudes of gradients along a range of position angles are displayed in Figure \ref{fig:fila_grad}(a), with a peak around the P.A. connecting IRS1 and HRF40, 
which we adopt as the direction of the filament.
The maximum gradient is 8.1 $\pm$ 0.8 km s$^{-1}$ pc$^{-1}$.
If we suppose the filament is a cylinder with uniform density, with material flowing towards IRS1, we can deduce a mass inflow rate $\dot{M}$ as (e.g., \citet{kirk2013})
\begin{equation}\label{eqn:inf_r}
\left ( \frac{\dot{M}}{M_\odot\text{Myr}^{-1}}\right ) = 1.02
\left ( \frac{\nabla V}{\text{km s}^{-1}\text{ pc}^{-1}}\right )
\left ( \frac{M_{\text{fila}}}{M_\odot}\right )
\end{equation}
where $\nabla V$ is the velocity gradient. $M_\text{fila}$ is the mass of the filament, 
which can be estimated by calculating a column density enclosed in the filament, as
\begin{align}\label{eqn:mass}
\left (\frac{M_{\text{enc}}}{M_\odot} \right ) &\notag= 1.13\times10^{-4}\mu_\text{H$_2$}\\
&\left ( \frac{m_\text{H}}{\text{kg}}\right )
 \left ( \frac{D}{\text{pc}}\right )^2
 \left ( \frac{A}{\text{arcsec}^2}\right )
 \left ( \frac{N(\text{H$_2$})}{\text{cm}^{-2}}\right )
\end{align}
where the enclosed mass is denoted as $M_{\text{enc}}$.
$m_\text{H}$ is the atomic weight of hydrogen, $D$ is the distance ($=$250pc), 
and $A$ is the enclosed area. 
$N$ indicates the column density of H$_2$. 
The abundance ratio of H$_2$ and C$^{18}$O in the L1455 region is estimated to be $N(\text{H$_2$}) = 1 \times 10^{7} N(\text{C$^{18}$O})$ \citep{curtis20101,lee2014}. 
The mean molecular weight of gas, $\mu_\text{H$_2$}$, is 2.7 per H$_2$ \citep{nishimura2015}. 
Thus, 
the filament has a mass $M_\text{fila}= 0.34M_\odot$, 
where we have adopted an excitation temperature of 12.6K \citep{curtis20101}.

Figure \ref{fig:fila_grad}(b) illustrates the mass inflow rate (Equation \ref{eqn:inf_r}) against different P.A.
Along the filament, we find $\dot{M}=2.8$ M$_\odot$ Myr$^{-1}$. 
Since a velocity gradient is clearly detected over an extended spatial area,
the filament is unlikely in the plane-of-sky or aligned with the line-of-sight.
For a projection angle in the range between $30^\circ$ to $60^\circ$, 
the mass infalling rate is a few solar masses per Myr.
On the other hand, the accretion rate towards IRS1 can also be inferred from the bolometric luminosity of the protostar if the gravitational energy of the infalling material is fully converted into radiation when reaching the star. This is given by $ L_\text{bol} = GM_\ast\dot{M}/R_\ast$, where the subscripted asterisks denote quantities of the protostar, yielding 
\begin{align}
\dot{M} = 3.20\times10^{-2}
\left ( \frac{M_\ast}{M_\odot}\right )^{-1}
\left ( \frac{R_\ast}{R_\odot}\right )
\left ( \frac{L_\text{bol}}{L_\odot}\right ).
\end{align}
For $L_{bol}$=3.6$L_\odot$ \citep{dunham2013} and $R_\ast \approx$ 3--5$R_\odot$ \citep{stahler1980}, we have $\dot{M}\approx 1-2$ M$_\odot$ Myr$^{-1}$. 
This is on the same order as calculated for the large-scale filament. 
Therefore, the detected filament could be the reservoir and feeding source for the protostellar system, providing significant infalling material. 
Since the time scale of a typical Class $\mathrm{0}/\mathrm{I}$ protostars is $\sim 10^5 - 10^6$ Myr (e.g., \citet{andre2000}), and the mass of the dense core is 0.54M$_\odot$, this mechanism, if maintained, can provide a similar amount of mass through the Class $\mathrm{0}/\mathrm{I}$ stages. 

We can further estimate the kinetic energy of the infalling material as $E_\text{k} = M_\text{fila}V_\text{inf}^2/2$,
where $V_\text{inf}$ is the weighted mean infalling velocity of the filament with respect to the systemic velocity. 
For $V_\text{sys}=4.70$ km s$^{-1}$ (Section \ref{sec:disk}), 
$V_\text{inf} = 0.22$ km s$^{-1}$. 
On the other hand, assuming a constant specific angular momentum for the dense core in IRS1
and assuming the core to be an isothermal sphere, the rotational energy
is $E_\text{R} = 1/2I\omega^2$
where $\omega$ is the angular velocity of the rotating core and $I=2/9M_\text{IRS1}R^2$ is its moment of inertia,
where the mass of the dense core ($\equiv M_\text{IRS1}$) encompassing L1455 IRS1 is 0.54M$_\odot$.
The ratio of the two energies is then $E_\text{k} / E_\text{R} \sim 3.5$.
This indicates that the filament is dynamically not negligible, but can possibly have significant impact on the core rotation.
We note that $I \propto M_\text{IRS1}R^2$ can vary by a factor of a few for different mass distributions. 
In conclusion, the faster-than-average envelope rotation seen in L1455 IRS1 could be related to the feeding from the associated filament.
In return, this could play a crucial role in the evolution of IRS1,
by effectively adding extra angular momentum to its rotational motion.

\subsection{Sample Trends: Competing Energies}

Dust polarization observations in L1455 IRS1 reveal an alignment between magnetic (B) field and outflow axis on the core scale ($\sim$1pc), while the two axes become perpendicular on the envelope scale ($\sim$1000AU, \citet{matthews2009,hull2014}). 
MHD simulations show that such a change in field orientation can be due to the interplay between B-fields and gas motions (e.g., \citet{machida2006}). 
Despite that such a change of orientations is seen in several sources (e.g., \citet{hull2013,hull2014}), unambiguous observational evidence
to connect this to gas motions remains scarce (e.g., \citet{yen2015}).
Nevertheless, with the increasing resolution in the study of gas kinematics in Class 0, 0/I sources (e.g., \citet{choi2010,murillo2013,ohashi2014}), the relation between B-fields and dynamics in protostars can now be discussed more systematically.
To that purpose, we compile a sample of Class 0, 0/I protostars where sufficient data -- from large and small scales, and from magnetic field morphologies  -- are available. 
We find 8 sources that satisfy these criteria (Table \ref{tb:sample}). 
We choose 5 cores (L1455 IRS1, L1448 IRS2, L1448 IRS 3B, L1157 mm, L1527 IRS) which have both single-dish and interferometric data. 
Three addtional sources are selected that show either resolved disks (NGC1333 IRAS 4A and VLA 1623A) or a clear hint of the presence of a rotating disk (L1448 mm). All 8 sources have polarization observations.
In Section \ref{sec:br}, we exemplify our analysis on L1455 IRS1 and 
discuss possible implications on its disk formation.
Sample trends are presented in Section \ref{sec:sample}.

\begin{deluxetable*}{lcccccc}[!ht]
\tablecaption{Source Sample\label{tb:sample}}
\tabletypesize{\scriptsize}

\tablehead{ \colhead{Source} & \multicolumn{2}{c}{Protostellar Position} & \colhead{Distance} & \colhead{$L_\text{bol}$} & \colhead{$T_\text{bol}$} & \colhead{Reference}\\ \cline{2-3}
           \colhead{(Class)} & \multicolumn{2}{c}{(J2000)} & \colhead{(pc)} & \colhead{($L_\sun$)} & \colhead{(K)} & \colhead{}} 
\startdata
L1455 IRS1      & 03$^h$27$^m$39$^s$1 & 30$^\circ$13$^\prime 03\farcs0$ & 250 & 3.6 & 65 & 1,2 \\
(0)             &                     &                                 &     &     &    &     \\
L1448 IRS2      & 03$^h$25$^m$22$^s$4 & 30$^\circ$45$^\prime 13\farcs3$ & 250 & 2.1 & 53 & 2,3 \\
(0/I)           &                     &                                 &     &     &    &     \\
L1448 IRS 3B    & 03$^h$25$^m$36$^s$3 & 30$^\circ$45$^\prime 14\farcs9$ & 250 & 4.3 & 90 & 2,3 \\
(0/I)           &                     &                                 &     &     &    &     \\
L1157 mm        & 20$^h$39$^m$06$^s$2 & 68$^\circ$02$^\prime 16\farcs6$ & 250 & 4.1 & 35 & 2,3 \\
(0)             &                     &                                 &     &     &    &     \\
L1527 IRS       & 04$^h$39$^m$53$^s$9 & 26$^\circ$03$^\prime 09\farcs8$ & 140 & 2.8 & 56 & 2,4 \\
(0/I)           &                     &                                 &     &     &    &     \\
NGC1333 IRAS 4A & 03$^h$29$^m$10$^s$4 & 31$^\circ$13$^\prime 32\farcs5$ & 250 & 4.2 & 51 & 2,3 \\
(0)             &                     &                                 &     &     &    &     \\
L1448 mm        & 03$^h$25$^m$38$^s$9 & 30$^\circ$44$^\prime 05\farcs4$ & 250 & 4.4 & 69 & 2,3 \\
(0)             &                     &                                 &     &     &    &     \\
VLA 1623A       & 16$^h$26$^m$26$^s$4 & -24$^\circ$24$^\prime 30\farcs7$ & 120 & 1.1 & 10 & 5,6 \\
(0)             &                     &                                 &     &     &    &     \\
\enddata
\tablerefs{(1) This work; (2) \citet{yen2015}; (3) \citet{enoch2009}; (4) \citet{tobin2011}; (5) \citet{murillo2013}; (6) \citet{murillo20132}}
\end{deluxetable*}

\subsubsection{Magnetic Field vs. Rotation: Change of Field Orientation Exemplified on L1455 IRS1}\label{sec:br}

Based on ideal MHD equations under strict flux-freezing conditions, the 
B-field strength $B$ is found to be $\propto n^{2/3}$ \citep{mestel1966},
where $n$ is the gas volume density.
The B-field strength is estimated to be $\sim5$ mG in the low-mass star forming region NGC1333 IRAS 4A, where n(H$_2$)$=4.3\times10^7$ cm$^{-3}$ \citep{girart2006}. 
Adopting this as a typical value for Class 0, $\mathrm{0}/\mathrm{I}$ sources, 
we deduce the field strengths in IRS1 according to the power law relation:
\begin{align}\label{eqn:bf}
\left ( \frac{B}{ \text{$\mu$G}} \right )=  
2.6 \times 10^{-2} \left ( \frac{n(H_2)}{\text{cm}^{-3}} \right )^{2/3}.   
\end{align}

We note that with this scaling we are able to reproduce field strength values similar to Zeeman measurements \citep{troland2008} for the clouds ($\sim$0.5 pc) embedding NGC1333 IRAS 4A, L1455 IRS1, and L1448 IRS2, IRS 3B.
Likewise, the observational estimate on a $\sim$1000 AU scale in L1157 mm \citep{stephens2013} also shows agreement.
From Equation \ref{eqn:bf}, we can calculate the B-field energy $E_\text{B}=VB^2/8\pi$, where $V$ is the volume.
In the previous section, we have computed kinetic and rotational energies for the filament and the rotating core. 
Moreover, the rotational energy
of a Keplerian disk can be expressed as $E_{\text{R}_\text{d}} = 2j_\text{d}^2M_\text{d}/R_\text{d}^2$ assuming a uniformly distributed disk mass $M_\text{d}$, where $j_\text{d}$ is the specific angular momentum at the disk edge of radius $R_\text{d}$.
With this, we find that the ratio $E_\text{B} / E_\text{R}$ changes from 3--4 on the filament/core scale to $E_\text{B} / E_{\text{R}_\text{d}}\sim 0.17$ on the disk scale (Table \ref{tb:cp}).
Thus, the significance of the magnetic fields towards smaller scales evolves from dominant to minor. 
This hints that the rotational motion starts to be more important than the B-field in the innermost region where the disk is present.
This supports the scenario that initially parallel field lines are dragged and bent by the rotating disk, and aligned along the disk major axis.

\subsubsection{Sample Analysis: Changes from Core to Disk Scale}\label{sec:sample}
In this section, we inspect whether there are common trends among the 8 Class 0, $\mathrm{0}/\mathrm{I}$ sources in our sample. 
The limited available data allow us only to compare energies on the two characteristic scales: disk (infalling envelope) and core.
Additionally, since the role of gravity is ignored in the previous discussion, 
we proceed to also compare B-field and rotational energy with gravitational energy.
We assume uniform mass distributions on each scale and rigid-body rotation for cores.
The energies on the core scale are calculated in a shell with the outer radius being the core radius, and the inner one being much larger than the rotating disk.
This division ensures that inner spin-up motions as well as higher densities in the innermost regions do not mimic or overweigh any core dynamics at larger scales that are relevant in our comparison.
The rotational and gravitational energies for disk and core are, therefore, given by
\begin{align}
\left\{\begin{matrix}
E_{\text{R}_\text{d}} = &\frac{2j_\text{d}^2M_\text{d}}{R_\text{d}^2}=\frac{2GM_\ast M}{R}
\\ 
E_{\text{R}_\text{c}} = &\frac{1}{2}I\omega^2 = 
                         \frac{1}{2}\frac{M_\text{enc}(R_\text{c}^2+R_\text{m}^2)}{2}\frac{j_\text{c}^2}{R_\text{c}^4}
\end{matrix}\right.
\end{align}
where the subscripts 'd' and 'c' denote values for disk and cores, respectively, and 
$R_\text{m}$ is the inner radius of the shell, and 
\begin{align}
\left\{\begin{matrix}
E_{\text{G}_\text{d}} = &\frac{3}{2}\frac{GM_\ast M_\text{d}}{R_\text{d}}+
                         \frac{3}{5}\frac{GM_\text{d}^2}{R_\text{d}}
\\ 
E_{\text{G}_\text{c}} = &\frac{G(M_\text{m}+M_\ast)}{R_\text{c}}M_\text{enc}
\end{matrix}\right.
\end{align}
where the energy is added from both self-gravitating potential and the central protostar, and $M_\text{m}+M_\ast$ is the total mass encompassed by the inner shell radius (gas and star), 
for the total gravitational energy exerted on the shell by all the mass inside.
The B-field energies $E_B$ are computed in the same way as in the previous section.

\begin{figure*}
\centering
\includegraphics[width=\textwidth]{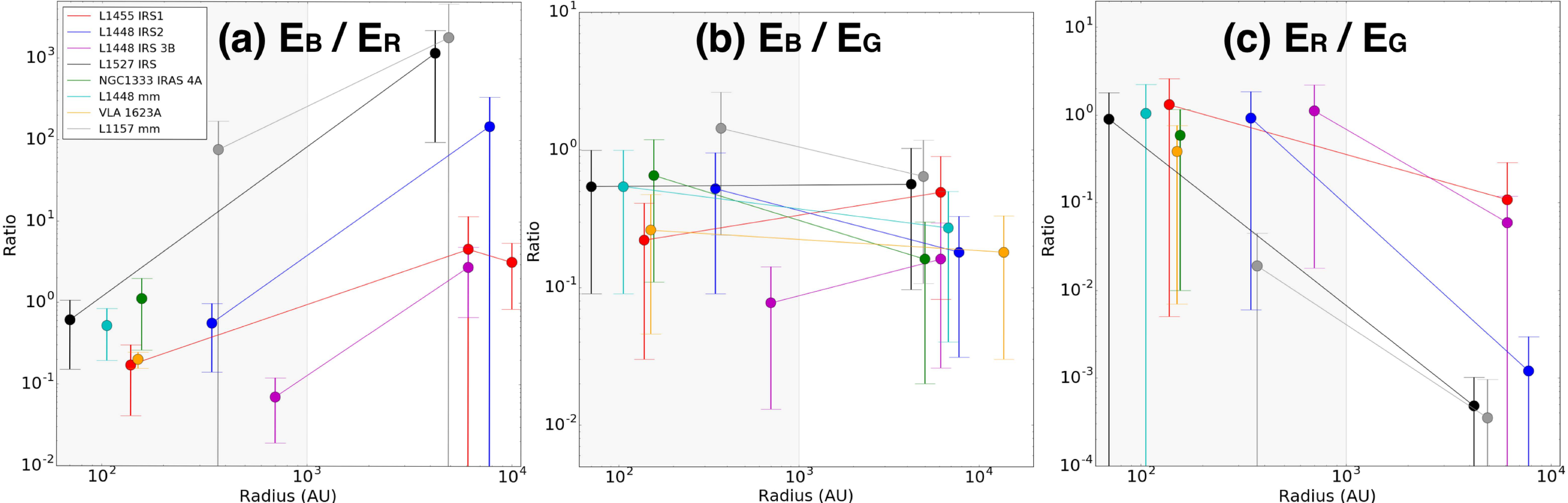}
\caption{Energy comparisons on disk and core scales in Class 0, $\mathrm{0}/\mathrm{I}$ sources. The gray-shaded region indicates the disk (envelope) scale ($<1000$ AU), and the right-half is the core scale. The error bars are obtained by adopting a factor of two uncertainties in $E_B$ and $M$ (e.g., $M\cdot 2^{\pm1}$), and assuming equal uncertainties in $M$ for all energies.}
\label{fig:ecp}
\end{figure*}

The results are summarized in Table~\ref{tb:cp}, Table~\ref{tb:cpdata} and illustrated in Figure \ref{fig:ecp}. Generally, the energy ratio
$E_B/E_R$ shows a decline from core to disk scale, consistent with what we find for IRS1, i.e., the 
rotational energy gains in significance from large to small scale, 
where it can be comparable or dominating over the B-field energy.
On the contrary, $E_B/E_G$ looks generally flat, with most values smaller than one.
This trend probably indicates that we are not yet tracing the even larger scales where the B-field could be dominating and preventing collapse.
Finally, we see an increasing importance of rotational energy as compared to gravitational energy with smaller scales. In particular, we note
that, on the disk (envelope) scale, $E_R/E_G\sim1$, which conforms with the expectation that the disks are rotationally supported and thus, prevent further infalling motions.

Finally, we note that, in our calculations we have assumed that B-field strengths scale following Equation \ref{eqn:bf}), cores follow a rigid-body rotation, and mass is uniformly distributed.
Although we have seen that the B-field strengths from independent observations give roughly the same values (Section \ref{sec:br}), the uncertainties can still be very large, and they are non-trivial to estimate.
Therefore, besides uncertainties given by fitting results of rotations (i.e., for $j$),
we include another factor of two uncertainties for $E_B$ and $M$, assuming that all the mass estimates are correlated, i.e., identical mass uncertainties apply to all energies.
They are depicted in Figure \ref{fig:ecp} as error bars.
While some sources show large error bars extending to zero, general trends are still apparent even in the presence of these large uncertainties.

\section{Summary and Conclusion}

We present new SMA and IRAM-30m single-dish observations toward the 
Class 0/I protostar L1455 IRS1 and its surrounding. Our main results are 
summarized in the following.

\begin{enumerate}

\item
{\it SMA High-Resolution Observations.} 
The SMA 1.3~mm continuum observation with a resolution of about $3''$ reveals
a compact dust component centered on IRS1 with a size around 300~AU. Its mass 
is estimated to be about $0.01 M_{\odot}$. With a spatial and velocity resolution around
 $4''$ and 0.069~km~s$^{-1}$ in C$^{18}$O (2--1), a clear velocity gradient of $\sim$150 km s$^{-1}$ pc$^{-1}$ is detected over a compact component of 1500~AU.
A clear outflow perpendicular to this gradient is seen in $^{12}$CO, extending out to about 6000~AU.
Additionally seen asymmetric features probably belong to a larger-scale envelope. 

\item
{\it IRAM-30m Larger-Scale Observation.}
With a resolution of about $12''$, the IRAM-30m observation in C$^{18}$O (2--1) over an 
area of about 0.2-by-0.2 pc captures the surroundings of IRS1 and partly also includes the 
protostars IRS4 and IRS5, and the starless cores HRF40 and CoreW.  A network of complex
structures, possibly bridges and filaments, in between these cores is apparent in the channel maps. IRS1 is found embedded in a dense core of about 0.05~pc in size with $0.35 M_{\odot}$.

\item
{\it Rotation and Disk in L1455 IRS1.}
The C$^{18}$O (2--1) emission towards IRS1 is optically thin, thus tracing the motion near the protostar. The lack of a clear gradient along the outflow direction in the P-V-diagram rules out 
significant infall motion. Fitting of the velocity gradient over the full range in the P-V-diagram perpendicular to the outflow direction is consistent with a rotational motion $v\sim r^p$ with 
a power-law index $p=-0.75$. Due to the limited instrumental resolution and the resulting blending 
of motions, this is likely indicative of a spin-up rotation, transitioning between an outer infalling and rotating motion with conserved angular momentum (observed with $p=-1$ when restricting the fitting to the largest scales, $\sim 300$ to 500~AU, in the P-V-diagram) and  the innermost detected scales
between 100 to 200~AU where the slope of the velocity profile flattens.
All this hints the likely presence of a Keplerian disk with a radius smaller than 200~AU and a
protostellar mass of $0.28 M_{\odot}$.

\item
{\it Core Rotation and Filament Connection.}
A velocity gradient of about 8 km s$^{-1}$ pc$^{-1}$ is found from the IRAM-30m C$^{18}$O (2--1)
emission over the core scale. Assuming the IRS1 core follows a rigid-body rotation, this leads to a 
specific angular momentum of about $5\times 10^{-3}$ km s$^{-1}$ pc at the core radius around 
5000~AU. This is about one order of magnitude larger than expected from a core-size-momentum scaling relation. 
Significant emission and structure is detected between the two cores IRS1 and 
HRF40. Probing a range of directions around IRS1 shows a maximum velocity gradient
of 8 km s$^{-1}$ pc$^{-1}$ precisely towards HRF40. Interpreting this as a gradient along the 
filamentary connection between the two cores yields a mass inflow of $\dot{M}=1.8$ M$_\odot$ Myr$^{-1}$, similar to the accretion rate onto IRS1 derived from its bolometric luminosity.
The resulting kinetic energy of the inflowing material is significant, a few times the core's rotational
energy. The filament is, thus, dynamically important, a main gas reservoir and possibly responsible 
for the larger-than-average core rotation.

\item
{\it Magnetic Field and Trends in Class 0/I Protostar Sample.}
In IRS1 the observed magnetic field orientation changes from parallel to the outflow axis on core scale to perpendicular on disk scale. This can be explained by our measured core and disk rotational 
energy which, compared to the magnetic field energy, grows from minor on large scale to 
dominating on small scale. The disk rotation is, thus, significant enough to bend the field lines
and align them with the disk velocity gradient. This growing importance of rotation on the disk 
scale is generally observed in our sample of 8 Class 0/I protostellar sources. Moreover, the
ratio of rotational-to-gravitational energy grows across the entire sample from large to small scale, approaching unity on disk scale which is consistent with the expectation of rotationally supported Keplerian disks. The magnetic-to-gravitational energy ratio remains roughly constant, smaller than one, over our tested scales.

\acknowledgments
We thank all the SMA and IRAM-30m staff supporting this work.
The SMA is a joint project between the Smithsonian Astrophysical
Observatory and the Academia Sinica Institute of Astronomy and
Astrophysics and is funded by the Smithsonian Institute and the
Academia Sinica.
PMK acknowledges support from the Ministry of Science and
Technology (MOST) of Taiwan (MOST 104-2119-M-001-019-MY3) and from an Academia Sinica Career Development Award.

\end{enumerate}

\begin{deluxetable*}{ccccccccc}[!ht]
\tablewidth{0pt}
\tabletypesize{\scriptsize}

\tablecaption{Energy Comparison in Class 0, 0/I Sources\label{tb:cp}}

\tablehead{
\colhead{Source}            & \colhead{Scale\tablenotemark{a}}& \colhead{$E_B$/$E_R$} &
\colhead{$E_B$/$E_G$}       & \colhead{$E_R$/$E_G$}       & 
\colhead{$\Delta\theta_\text{core}$} & \colhead{$\Delta\theta_\text{env}$}\\ 
\colhead{(Class)}           & \colhead{}                  & \colhead{} & 
\colhead{}                  & \colhead{}                  & 
\colhead{} & \colhead{} \\
\colhead{}                  & \colhead{}                  & \colhead{} & 
\colhead{}                  & \colhead{}                  & 
\colhead{} & \colhead{} \\
} 

\startdata
L1455 IRS1                  & disk                        & 0.17(0.13) &
0.22(0.19)                  & 1.3(1.3)                    & 6$^\circ$ & 84$^\circ$\\
($\mathrm{0}/\mathrm{I}$)   & core                        & 4.5(6.9) &
0.49(0.41)                  & 0.11(0.18)    && \\
                            & filament                    & 2.7(2.0) &
--                          & --            &&\\
\\
L1448 IRS2                  & disk                        & 0.56(0.42) &
0.52(0.43)                  & 0.92(0.91)                  & 15$^\circ$ & 3$^\circ$\\
($\mathrm{0}/\mathrm{I}$)   & core                        & 147(191) &
0.18(0.15)                  & 0.0012(0.0018)  &&\\
\\
L1448 IRS 3B                & disk                        & 0.069(0.050) &
0.077(0.064)                & 1.12(1.10)                  & 82$^\circ$ & 79$^\circ$\\
($\mathrm{0}/\mathrm{I}$)   & core                        & 2.7(2.1) &
0.16(0.13)                  & 0.059(0.059)  &&\\
\\
L1157 mm                    & envelope\tablenotemark{b}   & 76(94) &
1.4(1.2)                    & 0.019(0.026)                & 14$^\circ$ & 3$^\circ$\\
($\mathrm{0}$)              & core                        & 1800(2900)&
0.64(0.53)                  & 0.00035(0.00061) &&\\
\\
L1527 IRS                   & disk                        & 0.61(0.46) &
0.54(0.45)                  & 0.90(0.90)                  & 32$^\circ$ & 87$^\circ$\\
($\mathrm{0}/\mathrm{I}$)   & core                        & 1200(1100) &
0.56(0.46)                  & 0.00048(0.00054)&&\\
\\
NGC1333 IRAS 4A             & disk                        & 1.1(0.86) &
0.65(0.54)                  & 0.58(0.57)                  & 37$^\circ$ & $\sim90^\circ$\tablenotemark{c} \\
($\mathrm{0}$)              & core                        & -- &
0.16(0.14)                  & --            && \\
\\
L1448 mm                    & disk                        & 0.52(0.33) &
0.54(0.45)                  & 1.0(1.2)                    & 44$^\circ$ & 45$^\circ$\\
($\mathrm{0}$)              & core                        & -- &
0.27(0.23)                  & --            && \\
\\
VLA 1623A                   & disk                        & 0.20(0.05) &
0.26(0.21)                  & 1.30(1.29)                  & 60$^\circ$ & 83$^\circ$\\
($\mathrm{0}$)              & core                       & -- &
0.18(0.15)                  & --            &&
\enddata
\tablecomments{The values in parentheses are the uncertainties assuming a factor of two uncertainties in $E_B$ and $M$, and including errors from $j$ (see Table \ref{tb:cpdata}). The field orientations on core and envelope scale, $\Delta\theta_\text{core}$ and $\Delta\theta_\text{env}$, are the weighted averages from \citet{hull2014}}.
\tablenotetext{a}{The disk (or envelope) and core scales correspond to the smaller and larger radii in Table \ref{tb:cpdata}. The rotational energy for the filament scale in L1455 IRS1 is the kinetic energy.}
\tablenotetext{b}{Because $R \gg R_\text{d}$ in this source, we assume an infalling envelope with conserved angular momentum. See note (d) in Table \ref{tb:cpdata}}.
\tablenotetext{c}{The field orientation angle on envelope scale is estimated from polarization observed on a 100 AU scale in IRAS 4A1 \citep{cox2015}.}
\end{deluxetable*}

\appendix

\section{Combining SMA and IRAM-30m data: Reliability Test}\label{sec:test}

\begin{figure*}[!ht]
\centering
\includegraphics[width=\textwidth]{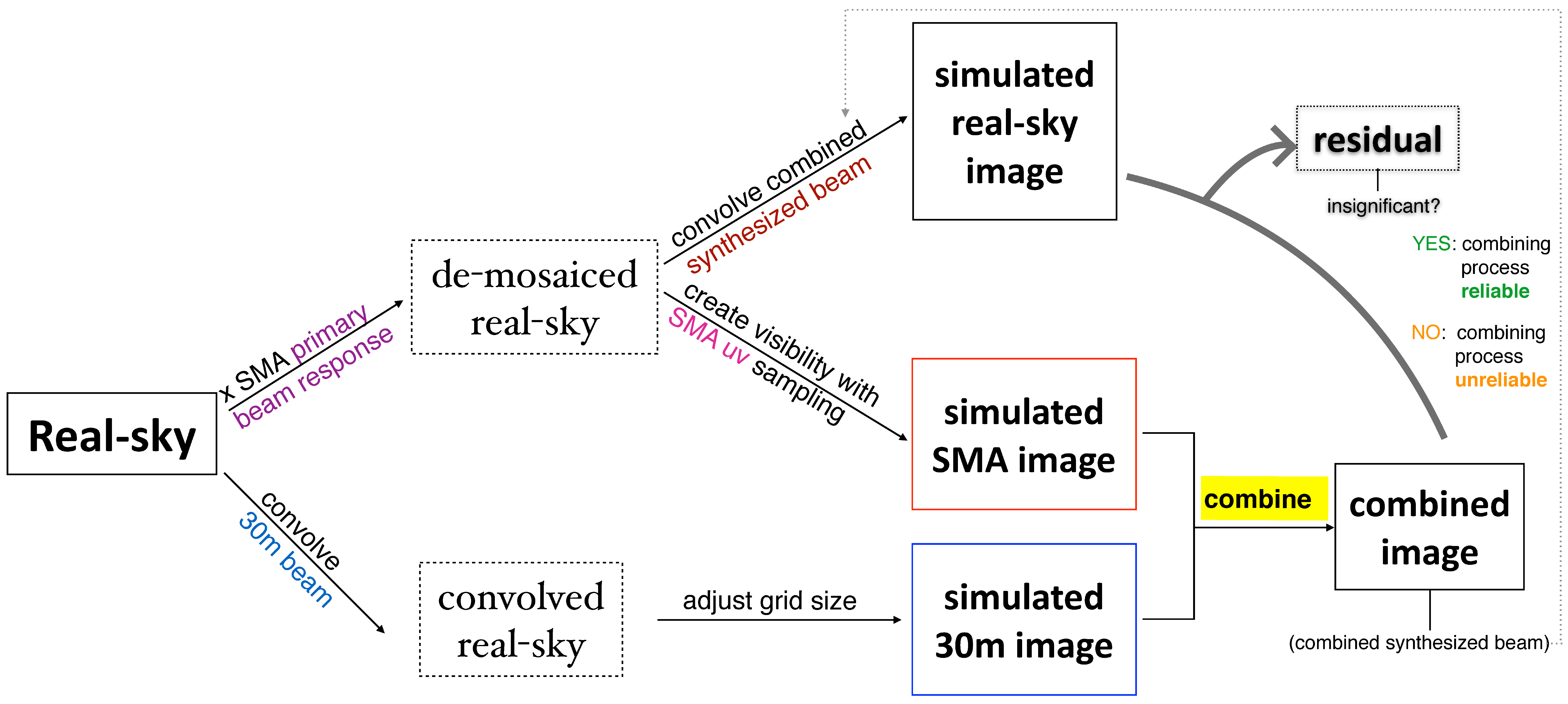}
\caption{Schematic flow chart to test combination process. An image with characteristics similar to our source is adopted as the real sky. The central black, red, and blue boxes denote the simulated observed images of the real sky. The bottom two are simulated SMA and single-dish images. The validity of the combination process is evaluated by the significance of the residual between combined and expected image.}
\label{fig:cbsim}
\end{figure*}

In order to check if the combination process by which we generate 
SMA-IRAM-combined maps does not remove real structures nor create artificial ones, we simulate the process as described below. The idea is illustrated in Figure \ref{fig:cbsim}. First, we choose a map to represent the real sky. The real sky is then converted into a simulated SMA and a simulated single-dish image (denoted simulated 30m image in Figure \ref{fig:cbsim}). After this, the two simulated images are combined through the above outlined method.
To assess the quality of this combined map, we generate another simulated image, which is the (simulated) observed map with a synthesized beam of  size equal to that of the combined map. Finally, this (simulated) observed map is compared with the combined map.

Since size and geometry of a source will generally affect what the source looks like in interferometric maps, we use the moment 0 of the combined map (Figure \ref{fig:m0}(b)) as the real sky. 
Furthermore, we are assured that the simulation does not produce non-ignorable artificial structures, which were present in \citet{yen2011} because of the gap between their single-dish and the SMA \textit{uv}-coverages. To simulate observations of the SMA, we multiply the real sky by the primary beam response of the SMA and convolve it with the SMA synthesized beam. For the single-dish IRAM-30m, the real sky is convolved with the 30m beam and then resampled onto the Nyquist grid as the real 30m map ($5\farcs602$). The combination of the two subsequently gives a "combined synthesized beam", which is used to simulate the observed map of the real sky (indicated by the dashed arrow in Figure \ref{fig:cbsim}). 

\begin{figure*}
\centering
\includegraphics[width=.99\textwidth]{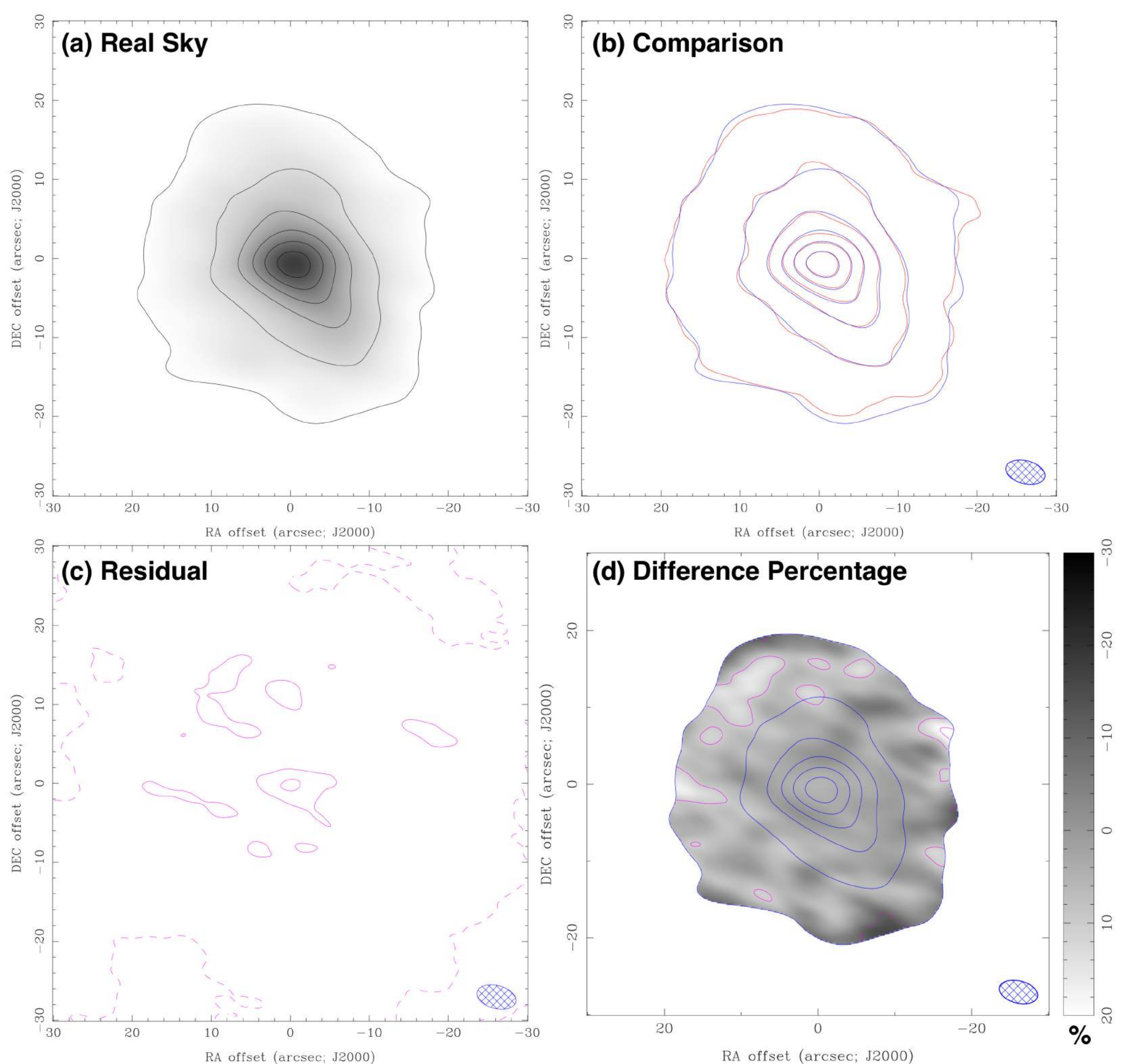}
\caption{(a) Real sky chosen for simulation. (b) Simulated observed image (blue contours) overlaid on combined map (red contours). Contour levels are from 3$\sigma$ in steps of 3$\sigma$ in (a) and (b). (c) Residual of combined map, subtracting observed one from (b). Contour levels are in steps of 0.5$\sigma$ of the observed map. Dashed lines indicate negative values. (d) Percentage with respect to the observed map. Overlaid are the observed contours (blue). The region with intensity lower than 3$\sigma$ of the observed map is masked (blank region). Contour levels are -10\%, 10\%, and 20\% respectively. A cross-hatched ellipse in the bottom right corner in each panel denotes the synthesized beam size.}
\label{fig:cbres}
\end{figure*}

Finally, the (simulated) observed map is subtracted from the combined map, giving an absolute residual map (Figure \ref{fig:cbres}(c)). Dividing this residual map by the observed map gives a map in unit of percentage, showing how significant the relative difference is (Figure \ref{fig:cbres}(d)). The results are shown in Figure \ref{fig:cbres}. Except for the inner most region, the combined map is consistent with the observed map within a 1$\sigma$ difference. In regions where the intensity is greater than 6$\sigma$, we find less than 10\% flux deviation.

In conclusion, the simulation assures that the combination of the SMA and the IRAM-30m maps is credible. We note that the best weighting for reproducing the best non-distorted image varies from case to case. Since \textit{uv}-samplings and beam sizes determine how well structures are recovered in combined maps, one should be careful when combining multiple data sets for different sources, with varying structures. We note, however, that it is impossible to perfectly reproduce the real-sky structures, due to the same reasons. We believe that a $<$10\% loss in the central region is acceptable, given that the SMA usually has the same level of uncertainty in flux.

\clearpage
\begin{landscape}
\begin{deluxetable*}{cccccccccccl}
\tablewidth{0pt}
\tabletypesize{\scriptsize}

\tablecaption{Properties of Class 0, 0/I Sources\label{tb:cpdata}}

\tablehead{
\colhead{Source}            & \colhead{$M_\ast$}            & \colhead{$R_\text{d}$} &
\colhead{R\tablenotemark{a}}& \colhead{$M_\text{enc}$}      & \colhead{N(H$_2$)} &
\colhead{$\mathbf{B}$}      & \colhead{$j$\tablenotemark{b}}& \colhead{$E_B$}    & 
\colhead{$E_R$}             & \colhead{$E_G$}               & \colhead{Ref.}            \\ 
\colhead{(Class)}           & \colhead{($M_\sun$)}        & \colhead{(AU)} &
\colhead{(AU)}              & \colhead{($M_\odot$)}       & \colhead{($\times10^5$)}    &
\colhead{(mG)}              & \colhead{($\times10^{-4}$)} &  \colhead{($\times 10^{40}$)} &
\colhead{($\times10^{40}$)} & \colhead{($\times10^{40}$)} & \colhead{}                  \\
\colhead{}                  & \colhead{}                  & \colhead{} & 
\colhead{}                  & \colhead{}                  & \colhead{(cm$^{-3})$}  &
\colhead{}                  & \colhead{(km s$^{-1}$ pc)}  & \colhead{(erg)} & 
\colhead{(erg)}             & \colhead{(erg)}             & \colhead{}                  \\
} 

\startdata
L1455 IRS1                  & 0.28                        & 180 &
140                         & 0.011                       & 700 &
7.0                         & 10(0.9)                     & 14  &
79(56)                      & 60(84)                      & 1,2,3,4\\
($\mathrm{0}/\mathrm{I}$)   &                             &     &
6125-4900                   & 0.35                        & 0.52&
0.06                        & 46(31)                      & 40  &
8.7(13)                     & 81(110)                     & \\
                            &                             &     &
10000\tablenotemark{c}      & 0.34                        & 0.81&
0.08                        & --                          & 44  &
16                          & --                          &\\
\\
L1448 IRS2                  & 0.18                        & 260 & 
340                         & 0.046                       & 190 & 
2.9                         & 8.7(0.7)                    & 37  & 
65(46)                      & 70(98)                      & 2,3,4,5\\
($\mathrm{0}/\mathrm{I}$)   &                             &     &
7750-4900                   & 0.56                        & 0.27  & 
0.04                        & 15(7.9)                     & 50  &
0.34(0.43)                  & 280(390)                    &\\
\\
L1448 IRS 3B                & 1.5                         & 530 &
700                         & 0.068                       & 33  & 
0.90                        & 38(1.5)                     & 30 &
440(310)                    & 390(540)                    & 2,3,4,5\\
($\mathrm{0}/\mathrm{I}$)   &                             &     & 
6125-4900                   & 3.7                         & 5.4 & 
0.27                        & 130(13)                     & 890 &
330(240)                    & 5500(7700)                  &\\
\\
L1157 mm                    & 0.05                        & $<$5 &
370\tablenotemark{d}        & 0.13                        & 430  & 
5.0                         & $<$0.5(0.25)                & 140  &
1.8(2.2)                    & 9.5(13)                     & 2,5,6\\
($\mathrm{0}$)              &                             &     &
4900-3500                   & 0.22                        & 0.49 &
0.05                        & 2.1(1.5)                    & 25  &
0.014(0.022)                & 39(54)                      &\\
\\
L1527 IRS                   & 0.16                        & 110 &
70                          & 0.036                       & 17000 & 
60                          & 5.2(0.6)                    & 130 &
210(160)                    & 240(330)                    & 2,5,6\\
($\mathrm{0}/\mathrm{I}$)   &                             &     &
4200-2750                   & 0.71                        & 2.3 &
0.16                        & 4.1(1.1)                    & 140 &
0.12(0.10)                  & 250(340)                    &\\
\\
NGC1333 IRAS 4A             & 0.14                        & 680 &
160                         & 0.56                        & 24000 &
74                          & 15(0.5)                     & 2300 &
2000(1400)                  & 3500(4800)                  & 2,3,4\\
($\mathrm{0}$)              &                             &     &  
5000\tablenotemark{e}       & 7.8                         & 10 & 
0.42                        & --                          & 2300 & 
--                          & 14000(20000)                & \\
\\
L1448 mm                    & 0.21                        & 150 &
110                         & 0.11                        & 15000 &
54                          & 7.9(2.3)                    & 380 &
730(660)                    & 700(970)                    & 2,3,4\\
($\mathrm{0}$)              &                             &     &  
6750\tablenotemark{e}       & 1.9                         & 1.0 & 
0.089                       & --                          & 260 & 
--                          & 960(1300)                   & \\
\\
VLA 1623A\tablenotemark{f}  & 0.22                        & 150 & 
150                         & 0.008                       & 390 & 
4.8                         & 8.3(--)                     & 8.1 & 
41(29)                      & 32(44)                      & 3,4,7\\
($\mathrm{0}$)              &                             &     &  
14000\tablenotemark{e}      & 0.24                        & 0.015 & 
0.005                       & --                          & 8.3 & 
--                          & 45(63)                      &
\enddata
\tablecomments{The values in parentheses are the uncertainties assuming a factor of two uncertainties in $E_B$ and $M$, and including errors from $j$. Error bars for disk $j$ are half of the best-fit results in \citet{yen2015}'s model. The radius is determined from the mean value of the FWHM size of disk or core. Estimation of $M$ and N(H$_2$) assumes the same abundance ratio as in L1455 IRS1. }
\tablenotetext{a}{On the core scale, $R$ stands for the outer and inner radius of the shell taken, respectively, where the outer one is the core size measured by single-dish surveys.}
\tablenotetext{b}{$j$ is the specific angular momentum measured at the disk edge for the small $R$ and from the rotational velocity of each core at the core radius for the large $R$.}
\tablenotetext{c}{Filament scale. $M_\text{enc}$ is the filament mass, and $E_R$ is replaced by $E_k$ of the filament.}
\tablenotetext{d}{Since $R \gg R_\text{d}$ in this source, we assume an infalling envelope with conserved angular momentum. Thus, $E_R = \frac{2Mj_\text{d}^2\ln(R/R_\text{d})}{R^2}$, integrated from the disk edge to $R$.}
\tablenotetext{e}{In NGC1333 IRAS 4A, L1448 mm, and VLA 1623A, the energies are calculated from the whole region (i.e., no shell division).}
\tablenotetext{f}{The envelope mass is calculated assuming the disk size for total flux observed in \citet{murillo2013}.}
\tablerefs{(1) This paper, (2) \citet{yen2015}, (3) \citet{enoch2009}, (4) \citet{enoch2006},
           (5) H.-W. Yen, (private communication), (6) \citet{motte2001}, (7) \citet{murillo2013}.}
\end{deluxetable*}
\clearpage
\end{landscape}


\begin{thebibliography}{}
\providecommand\natexlab[1]{#1}
\providecommand\JournalTitle[1]{#1}

\bibitem[{{Andr{\'e}} {et~al.}(2000){Andr{\'e}}, {Ward-Thompson}, \&
  {Barsony}}]{andre2000}
{Andr{\'e}}, P., {Ward-Thompson}, D., \& {Barsony}, M. 2000,
  \href{http://adsabs.harvard.edu/abs/2000prpl.conf...59A}{\JournalTitle{Protostars
  and Planets IV}, 59}

\bibitem[{{Andr{\'e}} {et~al.}(2010){Andr{\'e}}, {Men'shchikov}, {Bontemps},
  {K{\"o}nyves}, {Motte}, {Schneider}, {Didelon}, {Minier}, {Saraceno},
  {Ward-Thompson}, {di Francesco}, {White}, {Molinari}, {Testi}, {Abergel},
  {Griffin}, {Henning}, {Royer}, {Mer{\'{\i}}n}, {Vavrek}, {Attard},
  {Arzoumanian}, {Wilson}, {Ade}, {Aussel}, {Baluteau}, {Benedettini},
  {Bernard}, {Blommaert}, {Cambr{\'e}sy}, {Cox}, {di Giorgio}, {Hargrave},
  {Hennemann}, {Huang}, {Kirk}, {Krause}, {Launhardt}, {Leeks}, {Le Pennec},
  {Li}, {Martin}, {Maury}, {Olofsson}, {Omont}, {Peretto}, {Pezzuto}, {Prusti},
  {Roussel}, {Russeil}, {Sauvage}, {Sibthorpe}, {Sicilia-Aguilar}, {Spinoglio},
  {Waelkens}, {Woodcraft}, \& {Zavagno}}]{andre2010}
{Andr{\'e}}, P., {Men'shchikov}, A., {Bontemps}, S., {et~al.} 2010,
  \href{http://dx.doi.org/10.1051/0004-6361/201014666}{\JournalTitle{\aap},
  518, L102}

\bibitem[{{Anglada} {et~al.}(1989){Anglada}, {Rodriguez}, {Torrelles},
  {Estalella}, {Ho}, {Canto}, {Lopez}, \& {Verdes-Montenegro}}]{anglada1989}
{Anglada}, G., {Rodriguez}, L.~F., {Torrelles}, J.~M., {et~al.} 1989,
  \href{http://dx.doi.org/10.1086/167486}{\JournalTitle{\apj}, 341, 208}

\bibitem[{{Arce} {et~al.}(2010){Arce}, {Borkin}, {Goodman}, {Pineda}, \&
  {Halle}}]{arce2010}
{Arce}, H.~G., {Borkin}, M.~A., {Goodman}, A.~A., {Pineda}, J.~E., \& {Halle},
  M.~W. 2010,
  \href{http://dx.doi.org/10.1088/0004-637X/715/2/1170}{\JournalTitle{\apj},
  715, 1170}

\bibitem[{{Baars} {et~al.}(1987){Baars}, {Hooghoudt}, {Mezger}, \& {de
  Jonge}}]{iram}
{Baars}, J.~W.~M., {Hooghoudt}, B.~G., {Mezger}, P.~G., \& {de Jonge}, M.~J.
  1987, \JournalTitle{\aap}, 175, 319

\bibitem[{{Bachiller} \& {Cernicharo}(1986)}]{bachiller19861}
{Bachiller}, R., \& {Cernicharo}, J. 1986,
  \href{http://adsabs.harvard.edu/abs/1986A%26A...166..283B}{\JournalTitle{\aap},
  166, 283}

\bibitem[{{Bally} {et~al.}(1997){Bally}, {Devine}, {Alten}, \&
  {Sutherland}}]{bally1997}
{Bally}, J., {Devine}, D., {Alten}, V., \& {Sutherland}, R.~S. 1997,
  \href{http://adsabs.harvard.edu/abs/1997ApJ...478..603B}{\JournalTitle{\apj},
  478, 603}

\bibitem[{{Bally} {et~al.}(2008){Bally}, {Walawender}, {Johnstone}, {Kirk}, \&
  {Goodman}}]{bally2008}
{Bally}, J., {Walawender}, J., {Johnstone}, D., {Kirk}, H., \& {Goodman}, A.
  2008, {The Perseus Cloud}, ed. B.~{Reipurth}, 308

\bibitem[{{Beckwith} {et~al.}(1990){Beckwith}, {Sargent}, {Chini}, \&
  {Guesten}}]{beckwith1990}
{Beckwith}, S.~V.~W., {Sargent}, A.~I., {Chini}, R.~S., \& {Guesten}, R. 1990,
  \href{http://dx.doi.org/10.1086/115385}{\JournalTitle{\aj}, 99, 924}

\bibitem[{{Belloche}(2013)}]{belloche2013}
{Belloche}, A. 2013, \href{http://dx.doi.org/10.1051/eas/1362002}{in EAS
  Publications Series, Vol.~62, EAS Publications Series, ed. P.~{Hennebelle} \&
  C.~{Charbonnel}}, 25

\bibitem[{{Bergin} \& {Tafalla}(2007)}]{bergin2007}
{Bergin}, E.~A., \& {Tafalla}, M. 2007,
  \href{http://dx.doi.org/10.1146/annurev.astro.45.071206.100404}{\JournalTitle{\araa},
  45, 339}

\bibitem[{{\u{C}ernis}(1993)}]{cernis1993}
{\u{C}ernis}, K. 1993, \JournalTitle{Baltic Astronomy}, 2, 214

\bibitem[{{Choi} {et~al.}(2010){Choi}, {Tatematsu}, \& {Kang}}]{choi2010}
{Choi}, M., {Tatematsu}, K., \& {Kang}, M. 2010,
  \href{http://dx.doi.org/10.1088/2041-8205/723/1/L34}{\JournalTitle{\apjl},
  723, L34}

\bibitem[{{Cohen}(1980)}]{cohen1980}
{Cohen}, M. 1980, \href{http://dx.doi.org/10.1086/112630}{\JournalTitle{\aj},
  85, 29}

\bibitem[{{Cox} {et~al.}(2015){Cox}, {Harris}, {Looney}, {Segura-Cox}, {Tobin},
  {Li}, {Tychoniec}, {Chandler}, {Dunham}, {Kratter}, {Melis}, {Perez}, \&
  {Sadavoy}}]{cox2015}
{Cox}, E.~G., {Harris}, R.~J., {Looney}, L.~W., {et~al.} 2015,
  \href{http://dx.doi.org/10.1088/2041-8205/814/2/L28}{\JournalTitle{\apjl},
  814, L28}

\bibitem[{{Curtis} {et~al.}(2010{\natexlab{a}}){Curtis}, {Richer}, \&
  {Buckle}}]{curtis20101}
{Curtis}, E.~I., {Richer}, J.~S., \& {Buckle}, J.~V. 2010{\natexlab{a}},
  \href{http://dx.doi.org/10.1111/j.1365-2966.2009.15658.x}{\JournalTitle{\mnras},
  401, 455}

\bibitem[{{Curtis} {et~al.}(2010{\natexlab{b}}){Curtis}, {Richer}, {Swift}, \&
  {Williams}}]{curtis20102}
{Curtis}, E.~I., {Richer}, J.~S., {Swift}, J.~J., \& {Williams}, J.~P.
  2010{\natexlab{b}},
  \href{http://dx.doi.org/10.1111/j.1365-2966.2010.17214.x}{\JournalTitle{\mnras},
  408, 1516}

\bibitem[{{Davis} {et~al.}(1997{\natexlab{a}}){Davis}, {Eisloeffel}, {Ray}, \&
  {Jenness}}]{davis19972}
{Davis}, C.~J., {Eisloeffel}, J., {Ray}, T.~P., \& {Jenness}, T.
  1997{\natexlab{a}},
  \href{http://adsabs.harvard.edu/abs/1997A%26A...324.1013D}{\JournalTitle{\aap},
  324, 1013}

\bibitem[{{Davis} {et~al.}(1997{\natexlab{b}}){Davis}, {Ray}, {Eisloeffel}, \&
  {Corcoran}}]{davis19971}
{Davis}, C.~J., {Ray}, T.~P., {Eisloeffel}, J., \& {Corcoran}, D.
  1997{\natexlab{b}}, \JournalTitle{\aap}, 324, 263

\bibitem[{{Davis} {et~al.}(2008){Davis}, {Scholz}, {Lucas}, {Smith}, \&
  {Adamson}}]{davis2008}
{Davis}, C.~J., {Scholz}, P., {Lucas}, P., {Smith}, M.~D., \& {Adamson}, A.
  2008,
  \href{http://dx.doi.org/10.1111/j.1365-2966.2008.13247.x}{\JournalTitle{\mnras},
  387, 954}

\bibitem[{{de Zeeuw} {et~al.}(1999){de Zeeuw}, {Hoogerwerf}, {de Bruijne},
  {Brown}, \& {Blaauw}}]{de1999}
{de Zeeuw}, P.~T., {Hoogerwerf}, R., {de Bruijne}, J.~H.~J., {Brown}, A.~G.~A.,
  \& {Blaauw}, A. 1999,
  \href{http://dx.doi.org/10.1086/300682}{\JournalTitle{\aj}, 117, 354}

\bibitem[{{Di Francesco}(2012)}]{di2012}
{Di Francesco}, J. 2012,
  \href{http://adsabs.harvard.edu/abs/2012ASInC...4...13J}{in Astronomical
  Society of India Conference Series, Vol.~4, Astronomical Society of India
  Conference Series}, 13

\bibitem[{{Di Francesco} {et~al.}(2007){Di Francesco}, {Evans}, {Caselli},
  {Myers}, {Shirley}, {Aikawa}, \& {Tafalla}}]{di2007}
{Di Francesco}, J., {Evans}, II, N.~J., {Caselli}, P., {et~al.} 2007,
  \href{http://adsabs.harvard.edu/abs/2007prpl.conf...17D}{\JournalTitle{Protostars
  and Planets V}, 17}

\bibitem[{{Dunham} {et~al.}(2014){Dunham}, {Arce}, {Mardones}, {Lee},
  {Matthews}, {Stutz}, \& {Williams}}]{dunham2014}
{Dunham}, M.~M., {Arce}, H.~G., {Mardones}, D., {et~al.} 2014,
  \href{http://dx.doi.org/10.1088/0004-637X/783/1/29}{\JournalTitle{\apj}, 783,
  29}

\bibitem[{{Dunham} {et~al.}(2013){Dunham}, {Arce}, {Allen}, {Evans},
  {Broekhoven-Fiene}, {Chapman}, {Cieza}, {Gutermuth}, {Harvey}, {Hatchell},
  {Huard}, {Kirk}, {Matthews}, {Mer{\'{\i}}n}, {Miller}, {Peterson}, \&
  {Spezzi}}]{dunham2013}
{Dunham}, M.~M., {Arce}, H.~G., {Allen}, L.~E., {et~al.} 2013,
  \href{http://dx.doi.org/10.1088/0004-6256/145/4/94}{\JournalTitle{\aj}, 145,
  94}

\bibitem[{{Enoch} {et~al.}(2009){Enoch}, {Evans}, {Sargent}, \&
  {Glenn}}]{enoch2009}
{Enoch}, M.~L., {Evans}, II, N.~J., {Sargent}, A.~I., \& {Glenn}, J. 2009,
  \href{http://dx.doi.org/10.1088/0004-637X/692/2/973}{\JournalTitle{\apj},
  692, 973}

\bibitem[{{Enoch} {et~al.}(2006){Enoch}, {Young}, {Glenn}, {Evans}, {Golwala},
  {Sargent}, {Harvey}, {Aguirre}, {Goldin}, {Haig}, {Huard}, {Lange},
  {Laurent}, {Maloney}, {Mauskopf}, {Rossinot}, \& {Sayers}}]{enoch2006}
{Enoch}, M.~L., {Young}, K.~E., {Glenn}, J., {et~al.} 2006,
  \href{http://dx.doi.org/10.1086/498678}{\JournalTitle{\apj}, 638, 293}

\bibitem[{{Evans} {et~al.}(2009){Evans}, Dunham, J{\o}rgensen, Enoch, Mer?n,
  van Dishoeck, Alcal?, Myers, Stapelfeldt, Huard, Allen, Harvey, van Kempen,
  Blake, Koerner, Mundy, Padgett, \& Sargent}]{evans2009}
{Evans}, N. J.~I., Dunham, M.~M., J{\o}rgensen, J.~K., {et~al.} 2009,
  \href{http://stacks.iop.org/0067-0049/181/i=2/a=321}{\JournalTitle{The
  Astrophysical Journal Supplement Series}, 181, 321}

\bibitem[{{Frerking} {et~al.}(1982){Frerking}, {Langer}, \&
  {Wilson}}]{frerking1982}
{Frerking}, M.~A., {Langer}, W.~D., \& {Wilson}, R.~W. 1982,
  \href{http://dx.doi.org/10.1086/160451}{\JournalTitle{\apj}, 262, 590}

\bibitem[{{Friesen} {et~al.}(2013){Friesen}, {Kirk}, \&
  {Shirley}}]{friesen2013}
{Friesen}, R.~K., {Kirk}, H.~M., \& {Shirley}, Y.~L. 2013,
  \href{http://dx.doi.org/10.1088/0004-637X/765/1/59}{\JournalTitle{\apj}, 765,
  59}

\bibitem[{{Girart} {et~al.}(2006){Girart}, {Rao}, \& {Marrone}}]{girart2006}
{Girart}, J.~M., {Rao}, R., \& {Marrone}, D.~P. 2006,
  \href{http://dx.doi.org/10.1126/science.1129093}{\JournalTitle{Science}, 313,
  812}

\bibitem[{{Goldsmith} {et~al.}(1984){Goldsmith}, {Snell}, {Hemeon-Heyer}, \&
  {Langer}}]{goldsmith1984}
{Goldsmith}, P.~F., {Snell}, R.~L., {Hemeon-Heyer}, M., \& {Langer}, W.~D.
  1984, \href{http://dx.doi.org/10.1086/162635}{\JournalTitle{\apj}, 286, 599}

\bibitem[{{Goodman} {et~al.}(1993){Goodman}, {Benson}, {Fuller}, \&
  {Myers}}]{goodman1993}
{Goodman}, A.~A., {Benson}, P.~J., {Fuller}, G.~A., \& {Myers}, P.~C. 1993,
  \href{http://dx.doi.org/10.1086/172465}{\JournalTitle{\apj}, 406, 528}

\bibitem[{{Hacar} \& {Tafalla}(2011)}]{hacar2011}
{Hacar}, A., \& {Tafalla}, M. 2011,
  \href{http://dx.doi.org/10.1051/0004-6361/201117039}{\JournalTitle{\aap},
  533, A34}

\bibitem[{{Hatchell} {et~al.}(2007){Hatchell}, {Fuller}, {Richer}, {Harries},
  \& {Ladd}}]{hatchell20071}
{Hatchell}, J., {Fuller}, G.~A., {Richer}, J.~S., {Harries}, T.~J., \& {Ladd},
  E.~F. 2007,
  \href{http://dx.doi.org/10.1051/0004-6361:20066466}{\JournalTitle{\aap}, 468,
  1009}

\bibitem[{{Hatchell} {et~al.}(2005){Hatchell}, {Richer}, {Fuller},
  {Qualtrough}, {Ladd}, \& {Chandler}}]{hatchell2005}
{Hatchell}, J., {Richer}, J.~S., {Fuller}, G.~A., {et~al.} 2005,
  \href{http://dx.doi.org/10.1051/0004-6361:20041836}{\JournalTitle{\aap}, 440,
  151}

\bibitem[{{Helou} \& {Walker}(1988)}]{IRAS}
{Helou}, G., \& {Walker}, D.~W., eds. 1988, {Infrared astronomical satellite
  (IRAS) catalogs and atlases. Volume 7: The small scale structure catalog},
  Vol.~7

\bibitem[{{Ho} {et~al.}(2004){Ho}, {Moran}, \& {Lo}}]{sma}
{Ho}, P.~T.~P., {Moran}, J.~M., \& {Lo}, K.~Y. 2004,
  \href{http://dx.doi.org/10.1086/423245}{\JournalTitle{\apjl}, 616, L1}

\bibitem[{{Hull} {et~al.}(2013){Hull}, {Plambeck}, {Bolatto}, {Bower},
  {Carpenter}, {Crutcher}, {Fiege}, {Franzmann}, {Hakobian}, {Heiles}, {Houde},
  {Hughes}, {Jameson}, {Kwon}, {Lamb}, {Looney}, {Matthews}, {Mundy}, {Pillai},
  {Pound}, {Stephens}, {Tobin}, {Vaillancourt}, {Volgenau}, \&
  {Wright}}]{hull2013}
{Hull}, C.~L.~H., {Plambeck}, R.~L., {Bolatto}, A.~D., {et~al.} 2013,
  \href{http://dx.doi.org/10.1088/0004-637X/768/2/159}{\JournalTitle{\apj},
  768, 159}

\bibitem[{{Hull} {et~al.}(2014){Hull}, {Plambeck}, {Kwon}, {Bower},
  {Carpenter}, {Crutcher}, {Fiege}, {Franzmann}, {Hakobian}, {Heiles}, {Houde},
  {Hughes}, {Lamb}, {Looney}, {Marrone}, {Matthews}, {Pillai}, {Pound},
  {Rahman}, {Sandell}, {Stephens}, {Tobin}, {Vaillancourt}, {Volgenau}, \&
  {Wright}}]{hull2014}
{Hull}, C.~L.~H., {Plambeck}, R.~L., {Kwon}, W., {et~al.} 2014,
  \href{http://dx.doi.org/10.1088/0067-0049/213/1/13}{\JournalTitle{\apjs},
  213, 13}

\bibitem[{{Johnstone} {et~al.}(2004){Johnstone}, {Di Francesco}, \&
  {Kirk}}]{johnstone2004}
{Johnstone}, D., {Di Francesco}, J., \& {Kirk}, H. 2004,
  \href{http://dx.doi.org/10.1086/423737}{\JournalTitle{\apjl}, 611, L45}

\bibitem[{{J{\o}rgensen} {et~al.}(2008){J{\o}rgensen}, {Johnstone}, {Kirk},
  {Myers}, {Allen}, \& {Shirley}}]{jorgensen2008}
{J{\o}rgensen}, J.~K., {Johnstone}, D., {Kirk}, H., {et~al.} 2008,
  \href{http://dx.doi.org/10.1086/589956}{\JournalTitle{\apj}, 683, 822}

\bibitem[{{J{\o}rgensen} {et~al.}(2006){J{\o}rgensen}, {Harvey}, {Evans},
  {Huard}, {Allen}, {Porras}, {Blake}, {Bourke}, {Chapman}, {Cieza}, {Koerner},
  {Lai}, {Mundy}, {Myers}, {Padgett}, {Rebull}, {Sargent}, {Spiesman},
  {Stapelfeldt}, {van Dishoeck}, {Wahhaj}, \& {Young}}]{jorgensen2006}
{J{\o}rgensen}, J.~K., {Harvey}, P.~M., {Evans}, II, N.~J., {et~al.} 2006,
  \href{http://dx.doi.org/10.1086/504373}{\JournalTitle{\apj}, 645, 1246}

\bibitem[{{J{\o}rgensen} {et~al.}(2007){J{\o}rgensen}, {Bourke}, {Myers}, {Di
  Francesco}, {van Dishoeck}, {Lee}, {Ohashi}, {Sch{\"o}ier}, {Takakuwa},
  {Wilner}, \& {Zhang}}]{jorgensen20072}
{J{\o}rgensen}, J.~K., {Bourke}, T.~L., {Myers}, P.~C., {et~al.} 2007,
  \href{http://dx.doi.org/10.1086/512230}{\JournalTitle{\apj}, 659, 479}

\bibitem[{{Juan} {et~al.}(1993){Juan}, {Bachiller}, {Koempe}, \&
  {Martin-Pintado}}]{juan1993}
{Juan}, J., {Bachiller}, R., {Koempe}, C., \& {Martin-Pintado}, J. 1993,
  \href{http://adsabs.harvard.edu/abs/1993A%26A...270..432J}{\JournalTitle{\aap},
  270, 432}

\bibitem[{{Kirk} {et~al.}(2007){Kirk}, {Johnstone}, \& {Tafalla}}]{kirk2007}
{Kirk}, H., {Johnstone}, D., \& {Tafalla}, M. 2007,
  \href{http://dx.doi.org/10.1086/521395}{\JournalTitle{\apj}, 668, 1042}

\bibitem[{Kirk {et~al.}(2013)Kirk, Myers, Bourke, Gutermuth, Hedden, \&
  Wilson}]{kirk2013}
Kirk, H., Myers, P.~C., Bourke, T.~L., {et~al.} 2013,
  \href{http://stacks.iop.org/0004-637X/766/i=2/a=115}{\JournalTitle{The
  Astrophysical Journal}, 766, 115}

\bibitem[{{Kirk} {et~al.}(2005){Kirk}, {Ward-Thompson}, \&
  {Andr{\'e}}}]{kirk2005}
{Kirk}, J.~M., {Ward-Thompson}, D., \& {Andr{\'e}}, P. 2005,
  \href{http://dx.doi.org/10.1111/j.1365-2966.2005.09145.x}{\JournalTitle{\mnras},
  360, 1506}

\bibitem[{{Lee}(2010)}]{lee2010}
{Lee}, C.-F. 2010,
  \href{http://dx.doi.org/10.1088/0004-637X/725/1/712}{\JournalTitle{\apj},
  725, 712}

\bibitem[{{Lee} {et~al.}(2014){Lee}, {Stanimirovi{\'c}}, {Wolfire}, {Shetty},
  {Glover}, {Molina}, \& {Klessen}}]{lee2014}
{Lee}, M.-Y., {Stanimirovi{\'c}}, S., {Wolfire}, M.~G., {et~al.} 2014,
  \href{http://dx.doi.org/10.1088/0004-637X/784/1/80}{\JournalTitle{\apj}, 784,
  80}

\bibitem[{{Levreault}(1988)}]{levreault1988}
{Levreault}, R.~M. 1988,
  \href{http://dx.doi.org/10.1086/191275}{\JournalTitle{\apjs}, 67, 283}

\bibitem[{{Lynds}(1962)}]{lynds1962}
{Lynds}, B.~T. 1962,
  \href{http://dx.doi.org/10.1086/190072}{\JournalTitle{\apjs}, 7, 1}

\bibitem[{Machida {et~al.}(2006)Machida, Matsumoto, Hanawa, \&
  Tomisaka}]{machida2006}
Machida, M.~N., Matsumoto, T., Hanawa, T., \& Tomisaka, K. 2006,
  \href{http://stacks.iop.org/0004-637X/645/i=2/a=1227}{\JournalTitle{The
  Astrophysical Journal}, 645, 1227}

\bibitem[{{Matthews} {et~al.}(2009){Matthews}, {McPhee}, {Fissel}, \&
  {Curran}}]{matthews2009}
{Matthews}, B.~C., {McPhee}, C.~A., {Fissel}, L.~M., \& {Curran}, R.~L. 2009,
  \href{http://dx.doi.org/10.1088/0067-0049/182/1/143}{\JournalTitle{\apjs},
  182, 143}

\bibitem[{{McKee} \& {Ostriker}(2007)}]{MO2008}
{McKee}, C.~F., \& {Ostriker}, E.~C. 2007,
  \href{http://dx.doi.org/10.1146/annurev.astro.45.051806.110602}{\JournalTitle{\araa},
  45, 565}

\bibitem[{{Mestel}(1966)}]{mestel1966}
{Mestel}, L. 1966, \JournalTitle{\mnras}, 133, 265

\bibitem[{{Motte} \& {Andr{\'e}}(2001)}]{motte2001}
{Motte}, F., \& {Andr{\'e}}, P. 2001,
  \href{http://dx.doi.org/10.1051/0004-6361:20000072}{\JournalTitle{\aap}, 365,
  440}

\bibitem[{{Murillo} \& {Lai}(2013)}]{murillo20132}
{Murillo}, N.~M., \& {Lai}, S.-P. 2013,
  \href{http://dx.doi.org/10.1088/2041-8205/764/1/L15}{\JournalTitle{\apjl},
  764, L15}

\bibitem[{{Murillo} {et~al.}(2013){Murillo}, {Lai}, {Bruderer}, {Harsono}, \&
  {van Dishoeck}}]{murillo2013}
{Murillo}, N.~M., {Lai}, S.-P., {Bruderer}, S., {Harsono}, D., \& {van
  Dishoeck}, E.~F. 2013,
  \href{http://dx.doi.org/10.1051/0004-6361/201322537}{\JournalTitle{\aap},
  560, A103}

\bibitem[{{Nishimura} {et~al.}(2015){Nishimura}, {Tokuda}, {Kimura}, {Muraoka},
  {Maezawa}, {Ogawa}, {Dobashi}, {Shimoikura}, {Mizuno}, {Fukui}, \&
  {Onishi}}]{nishimura2015}
{Nishimura}, A., {Tokuda}, K., {Kimura}, K., {et~al.} 2015,
  \href{http://dx.doi.org/10.1088/0067-0049/216/1/18}{\JournalTitle{\apjs},
  216, 18}

\bibitem[{{Ohashi} {et~al.}(2014){Ohashi}, {Saigo}, {Aso}, {Aikawa},
  {Koyamatsu}, {Machida}, {Saito}, {Takahashi}, {Takakuwa}, {Tomida},
  {Tomisaka}, \& {Yen}}]{ohashi2014}
{Ohashi}, N., {Saigo}, K., {Aso}, Y., {et~al.} 2014,
  \href{http://dx.doi.org/10.1088/0004-637X/796/2/131}{\JournalTitle{\apj},
  796, 131}

\bibitem[{{Padoan} {et~al.}(1999){Padoan}, {Bally}, {Billawala}, {Juvela}, \&
  {Nordlund}}]{padoan1999}
{Padoan}, P., {Bally}, J., {Billawala}, Y., {Juvela}, M., \& {Nordlund}, {\AA}.
  1999, \href{http://dx.doi.org/10.1086/307864}{\JournalTitle{\apj}, 525, 318}

\bibitem[{{Pi{\'e}tu} {et~al.}(2007){Pi{\'e}tu}, {Dutrey}, \&
  {Guilloteau}}]{pietu2007}
{Pi{\'e}tu}, V., {Dutrey}, A., \& {Guilloteau}, S. 2007,
  \href{http://dx.doi.org/10.1051/0004-6361:20066537}{\JournalTitle{\aap}, 467,
  163}

\bibitem[{{Ridge} {et~al.}(2006){Ridge}, {Di Francesco}, {Kirk}, {Li},
  {Goodman}, {Alves}, {Arce}, {Borkin}, {Caselli}, {Foster}, {Heyer},
  {Johnstone}, {Kosslyn}, {Lombardi}, {Pineda}, {Schnee}, \&
  {Tafalla}}]{ridge2006}
{Ridge}, N.~A., {Di Francesco}, J., {Kirk}, H., {et~al.} 2006,
  \href{http://dx.doi.org/10.1086/503704}{\JournalTitle{\aj}, 131, 2921}

\bibitem[{{Sadavoy} {et~al.}(2010){Sadavoy}, {Di Francesco}, {Bontemps},
  {Megeath}, {Rebull}, {Allgaier}, {Carey}, {Gutermuth}, {Hora}, {Huard},
  {McCabe}, {Muzerolle}, {Noriega-Crespo}, {Padgett}, \&
  {Terebey}}]{sadavoy2010}
{Sadavoy}, S.~I., {Di Francesco}, J., {Bontemps}, S., {et~al.} 2010,
  \href{http://dx.doi.org/10.1088/0004-637X/710/2/1247}{\JournalTitle{\apj},
  710, 1247}

\bibitem[{{Sancisi} {et~al.}(1974){Sancisi}, {Goss}, {Anderson}, {Johansson},
  \& {Winnberg}}]{sancisi1974}
{Sancisi}, R., {Goss}, W.~M., {Anderson}, C., {Johansson}, L.~E.~B., \&
  {Winnberg}, A. 1974, \JournalTitle{\aap}, 35, 445

\bibitem[{{Sargent}(1979)}]{sargent1979}
{Sargent}, A.~I. 1979,
  \href{http://dx.doi.org/10.1086/157378}{\JournalTitle{\apj}, 233, 163}

\bibitem[{{Sault} {et~al.}(1995){Sault}, {Teuben}, \& {Wright}}]{miriad}
{Sault}, R.~J., {Teuben}, P.~J., \& {Wright}, M.~C.~H. 1995, in Astronomical
  Society of the Pacific Conference Series, Vol.~77, Astronomical Data Analysis
  Software and Systems IV, ed. R.~A. {Shaw}, H.~E. {Payne}, \& J.~J.~E.
  {Hayes}, 433

\bibitem[{{Scoville} {et~al.}(1993){Scoville}, {Carlstrom}, {Chandler},
  {Phillips}, {Scott}, {Tilanus}, \& {Wang}}]{mir}
{Scoville}, N.~Z., {Carlstrom}, J.~E., {Chandler}, C.~J., {et~al.} 1993,
  \href{http://dx.doi.org/10.1086/133332}{\JournalTitle{\pasp}, 105, 1482}

\bibitem[{{Sep{\'u}lveda} {et~al.}(2011){Sep{\'u}lveda}, {Anglada},
  {Estalella}, {L{\'o}pez}, {Girart}, \& {Yang}}]{sepulveda2011}
{Sep{\'u}lveda}, I., {Anglada}, G., {Estalella}, R., {et~al.} 2011,
  \href{http://dx.doi.org/10.1051/0004-6361/200912916}{\JournalTitle{\aap},
  527, A41}

\bibitem[{Simon(2009)}]{simon2009}
Simon, T. 2009,
  \href{http://stacks.iop.org/0004-637X/693/i=2/a=1803}{\JournalTitle{The
  Astrophysical Journal}, 693, 1803}

\bibitem[{{Stahler} {et~al.}(1980){Stahler}, {Shu}, \& {Taam}}]{stahler1980}
{Stahler}, S.~W., {Shu}, F.~H., \& {Taam}, R.~E. 1980,
  \href{http://dx.doi.org/10.1086/158377}{\JournalTitle{\apj}, 241, 637}

\bibitem[{{Stephens} {et~al.}(2013){Stephens}, {Looney}, {Kwon}, {Hull},
  {Plambeck}, {Crutcher}, {Chapman}, {Novak}, {Davidson}, {Vaillancourt},
  {Shinnaga}, \& {Matthews}}]{stephens2013}
{Stephens}, I.~W., {Looney}, L.~W., {Kwon}, W., {et~al.} 2013,
  \href{http://dx.doi.org/10.1088/2041-8205/769/1/L15}{\JournalTitle{\apjl},
  769, L15}

\bibitem[{{Sun} {et~al.}(2006){Sun}, {Kramer}, {Ossenkopf}, {Bensch},
  {Stutzki}, \& {Miller}}]{sun2006}
{Sun}, K., {Kramer}, C., {Ossenkopf}, V., {et~al.} 2006,
  \href{http://dx.doi.org/10.1051/0004-6361:20054256}{\JournalTitle{\aap}, 451,
  539}

\bibitem[{{Tafalla} \& {Hacar}(2015)}]{tafalla2015}
{Tafalla}, M., \& {Hacar}, A. 2015,
  \href{http://dx.doi.org/10.1051/0004-6361/201424576}{\JournalTitle{\aap},
  574, A104}

\bibitem[{{Tapia} {et~al.}(1997){Tapia}, {Persi}, {Bohigas}, \&
  {Ferrari-Toniolo}}]{tapia1997}
{Tapia}, M., {Persi}, P., {Bohigas}, J., \& {Ferrari-Toniolo}, M. 1997,
  \href{http://dx.doi.org/10.1086/118390}{\JournalTitle{\aj}, 113, 1769}

\bibitem[{{Tobin} {et~al.}(2011){Tobin}, {Hartmann}, {Chiang}, {Looney},
  {Bergin}, {Chandler}, {Masqu{\'e}}, {Maret}, \& {Heitsch}}]{tobin2011}
{Tobin}, J.~J., {Hartmann}, L., {Chiang}, H.-F., {et~al.} 2011,
  \href{http://dx.doi.org/10.1088/0004-637X/740/1/45}{\JournalTitle{\apj}, 740,
  45}

\bibitem[{{Troland} \& {Crutcher}(2008)}]{troland2008}
{Troland}, T.~H., \& {Crutcher}, R.~M. 2008,
  \href{http://dx.doi.org/10.1086/587546}{\JournalTitle{\apj}, 680, 457}

\bibitem[{{Ulrich}(1976)}]{ulrich1976}
{Ulrich}, R.~K. 1976,
  \href{http://dx.doi.org/10.1086/154840}{\JournalTitle{\apj}, 210, 377}

\bibitem[{{Ungerechts} \& {Thaddeus}(1985)}]{ungerechts1985}
{Ungerechts}, H., \& {Thaddeus}, P. 1985, in Bulletin of the American
  Astronomical Society, Vol.~17, Bulletin of the American Astronomical Society,
  607

\bibitem[{{Walker-Smith} {et~al.}(2014){Walker-Smith}, {Richer}, {Buckle},
  {Hatchell}, \& {Drabek-Maunder}}]{walker2014}
{Walker-Smith}, S.~L., {Richer}, J.~S., {Buckle}, J.~V., {Hatchell}, J., \&
  {Drabek-Maunder}, E. 2014,
  \href{http://dx.doi.org/10.1093/mnras/stu512}{\JournalTitle{\mnras}, 440,
  3568}

\bibitem[{{Ward-Thompson} {et~al.}(2007){Ward-Thompson}, {Andr{\'e}},
  {Crutcher}, {Johnstone}, {Onishi}, \& {Wilson}}]{ward2007}
{Ward-Thompson}, D., {Andr{\'e}}, P., {Crutcher}, R., {et~al.} 2007,
  \JournalTitle{Protostars and Planets V}, 33

\bibitem[{{Wilson} {et~al.}(1996){Wilson}, {Gaume}, {Johnston}, \&
  {Schmid-Burgk}}]{wilson1996}
{Wilson}, T.~L., {Gaume}, R.~A., {Johnston}, K.~J., \& {Schmid-Burgk}, J. 1996,
  in Science with Large Millimetre Arrays, ed. P.~A. {Shaver}, 177

\bibitem[{{Wu} {et~al.}(2007){Wu}, {Dunham}, {Evans}, {Bourke}, \&
  {Young}}]{wu2007}
{Wu}, J., {Dunham}, M.~M., {Evans}, II, N.~J., {Bourke}, T.~L., \& {Young},
  C.~H. 2007, \href{http://dx.doi.org/10.1086/511959}{\JournalTitle{\aj}, 133,
  1560}

\bibitem[{{Yen} {et~al.}(2015){Yen}, {Koch}, {Takakuwa}, {Ho}, {Ohashi}, \&
  {Tang}}]{yen2015}
{Yen}, H.-W., {Koch}, P.~M., {Takakuwa}, S., {et~al.} 2015,
  \href{http://dx.doi.org/10.1088/0004-637X/799/2/193}{\JournalTitle{\apj},
  799, 193}

\bibitem[{{Yen} {et~al.}(2011){Yen}, {Takakuwa}, \& {Ohashi}}]{yen2011}
{Yen}, H.-W., {Takakuwa}, S., \& {Ohashi}, N. 2011,
  \href{http://dx.doi.org/10.1088/0004-637X/742/1/57}{\JournalTitle{\apj}, 742,
  57}

\bibitem[{{Yen} {et~al.}(2013){Yen}, {Takakuwa}, {Ohashi}, \& {Ho}}]{yen2013}
{Yen}, H.-W., {Takakuwa}, S., {Ohashi}, N., \& {Ho}, P.~T.~P. 2013,
  \href{http://dx.doi.org/10.1088/0004-637X/772/1/22}{\JournalTitle{\apj}, 772,
  22}

\bibitem[{{Young} {et~al.}(2015){Young}, {Young}, {Lai}, {Dunham}, \&
  {Evans}}]{young2015}
{Young}, K.~E., {Young}, C.~H., {Lai}, S.-P., {Dunham}, M.~M., \& {Evans}, II,
  N.~J. 2015,
  \href{http://dx.doi.org/10.1088/0004-6256/150/2/40}{\JournalTitle{\aj}, 150,
  40}

\end{thebibliography}
\end{document}